\documentclass[11pt,superscriptaddress, onecolumn, nofootinbib]{revtex4-1}

\usepackage{amssymb,amsmath}
\usepackage{latexsym}
\usepackage{graphicx}
\usepackage{caption}
\usepackage{subcaption}
\usepackage{enumerate}
\usepackage{float}

\usepackage{color}
\usepackage{graphics}
\usepackage{epsfig}
\usepackage{bm}        

\begin{document}

\newcommand{\dR}{\mathbb R}
\newcommand{\dC}{\mathbb C}
\newcommand{\dZ}{\mathbb Z}
\newcommand{\id}{\mathbb I}
\newcommand{\ud}{\mathrm{d}}
\newcommand{\mfn}{\mathfrak{n}}

\title[]{Hamiltonian formalism and gauge-fixing conditions for cosmological perturbation theory}

\author{Przemys\l aw Ma\l kiewicz}

\address{National Centre for Nuclear Research, Pasteura 7, 02-093 Warsaw, Poland} \email{Przemyslaw.Malkiewicz@ncbj.gov.pl}

\begin{abstract}
We apply the Dirac procedure for constrained systems to the Arnowitt-Deser-Misner formalism linearized around the Friedmann-Lemaitre universe. We explain and employ some basic concepts such as Dirac observables, Dirac brackets, gauge-fixing conditions, reduced phase space, physical Hamiltonian and physical dynamics. In particular, we elaborate on the key concept which is the canonical isomorphism between different gauge-fixing surfaces. We apply our formalism to describe the reduced phase space of cosmological perturbations in some popular in the literature gauges. Our formalism is first developed for the universe with a single fluid and then extended to the multi-fluid case. The obtained results are a starting point for { complete} quantization of the cosmological perturbations and the cosmological background.  { Our approach may be used in future to derive the reduced phase space of higher order perturbations and in more generic cosmological spacetimes.} 
\end{abstract}

\maketitle

\section{Introduction}

We revisit the Hamiltonian formalism for cosmological perturbations (see in particular \cite{Anderegg:1994xq,Langlois:1994ec} for earlier treatments). { We apply some commonly used in canonical relativity concepts to establish the reduced phase space for a slightly perturbed universe filled with one or more perfect fluids. Despite the fact that the reduced phase space of cosmological perturbations has been derived before, our concepts and our derivation are new to cosmological perturbation theory and shed new light on it. Our approach may be applied to some standing problems such as deriving the reduced phase space for linear and higher-order perturbations in less symmetric cosmological backgrounds leading to their complete quantisation that includes both the cosmological background and the perturbations. We postpone a more detailed discussion of possible applications of our approach to Conclusions.} 

The basic method of our approach is the Dirac method for constrained systems combined with use of the canonical isomorphism between different gauge-fixing surfaces in the phase space and equipped with different Dirac brackets. Our starting point is the Arnowitt-Deser-Misner (ADM) formalism \cite{Arnowitt:1962hi} in which the canonical variables are the three-metric components and canonically conjugate momenta (encoding the extrinsic curvature of the three-geometry), $(q_{ab},\pi^{ab})$.  The ADM Hamiltonian is a sum of first-class constraints, in which the lapse function $N$ and the shift vector $N^i$ play the role of the Lagrange multipliers. The perturbative expansion of the ADM formalism includes the following features: 1) The zero-order Hamiltonian constraint $\mathcal{H}\big|^{(0)}\approx 0$ generates and constrains the physical dynamics of the background (zero-order) variables; 2) The second-order Hamiltonian $\mathcal{H}\big|^{(2)}$ generates the dynamics of the first-order perturbations, though, it is not a constraint. Nevertheless, its definition is ambiguous and includes four first-order first-class constraints $\mathcal{H}_0\big|^{(1)}\approx 0$ and $\mathcal{H}_i\big|^{(1)}\approx 0$ ($i=1,2,3$) multiplied by the first-order Lagrange multipliers $\delta N$ and $\delta N^i$. The non-vanishing term of the second-order Hamiltonian is $N\mathcal{H}_0\big|^{(2)}$, where $N$ is the zero-order lapse function.

{ The structure of the full gravitational Hamiltonian which is a sum of first-class constraints reflects the so-called ``time problem". The central issue is the fact that the dynamics of the gravitational field coincides with a gauge transformation and, as a corollary, gauge-invariant (or, physical) variables are constants of motion.  The time problem is present in the background dynamics that is generated by the constraint $\mathcal{H}\big|^{(0)}\approx 0$ (see Sec IV for some further comments). However, this is worth noting that in the perturbations' dynamics this problem does not arise as the second-order Hamiltonian includes the non-vanishing (i.e. non-constraining) term, $N\mathcal{H}_0\big|^{(2)}$.} A way to understand this is to notice that a genuine change of time by a first-order perturbation also produces second-order effects in the dynamics of linear perturbations. { Since the second-order effects are excluded from the present formalism, the first-order time-reparametrization invariance of the dynamics of linear perturbations is broken.}  

The classic reference on the Hamiltonian approach to cosmological perturbation theory is D. Langlois' paper \cite{Langlois:1994ec}, where the reduced phase space formalism for a universe minimally coupled to a scalar field in a potential is derived. The author employs a method derived in \cite{doi:10.1063/1.529065} and based on the Hamilton-Jacobi theory. This method is used to define a canonical transformation that separates physical, gauge-invariant degrees of freedom from gauge degrees of freedom in such a way that their canonical structures become also separate. As the author emphasizes, such a separation of variables allows to derive the physical Hamiltonian that generates the physical dynamics in a very efficient way since the gauge variables that make up in particular the complicated intrinsic curvature term in $\mathcal{H}\big|^{(2)}$ can be ignored. The Langlois approach is contrasted with the approach described in \cite{Mukhanov:1990me}, within which the derivation of the physical Hamiltonian involves significantly heavier computations.

Our approach employs the Dirac procedure \cite{Dirac:1964:LQM}. The Dirac procedure appears to be more developed and to deal with more aspects of constrained systems than the method employed in \cite{Langlois:1994ec}. Instead of making use of a canonical transformation of the mentioned type, the Dirac method starts with extending the set of constraints by introducing the gauge-fixing conditions, one for each first-class constraint, with purpose of obtaining a set of second-class constraints. Then the so-called Dirac brackets are introduced. This last step allows us to solve strongly all the constraints by reducing the number of canonical variables and to simplify the form of the Hamiltonian. In particular, we get rid of the first-order constraints  $\mathcal{H}_0\big|^{(1)}=0$ and $\mathcal{H}_i\big|^{(1)}=0$ from the Hamiltonian and the values of the Lagrange multipliers $\delta N$ and  $\delta N^i$ become irrelevant for defining the dynamics in the reduced phase space. Working in the spatially flat slicing gauge also removes the intrinsic curvature term (which becomes pure gauge) and we are able to derive the physical Hamiltonian as quickly as Langlois in his original paper. In order to reconstruct the full four-dimensional metric, one needs to determine $\delta N$ and  $\delta N^i$. This step is normally omitted in the Dirac analysis of constrained systems and is not very much discussed in \cite{Langlois:1994ec} either. As we shall see, one must study the consistency of the gauge-fixing conditions in the initial ADM phase space.

The main result of the paper is the derivation of the reduced phase spaces of cosmological perturbations together with respective reduced Hamiltonians via the Dirac procedure combined with use of the canonical isomorphism between different gauge-fixing surfaces equipped with inequivalent Dirac brackets. The Dirac procedure has never been used in the domain of cosmological perturbations so consistently and to such an extent. We show that there exists a unique and abstract physical phase space furnished with a unique and abstract physical Hamiltonian. Our derivation sheds more light on why the Langlois procedure leads to such a simple physical Hamiltonian which does not include the second-order curvature term. Furthermore, we show how different choices of gauge-fixing conditions in the phase space correspond to different physical interpretations of the gauge-invariant canonical variables. Thanks to the existence of the aforementioned canonical isomorphism, it is quite straightforward to obtain a canonical framework for all choices of gauge, once the framework is established for any single choice of gauge. Although, the corresponding properties are well-known to hold for the configuration space approach, our work derives and clarifies them at the level of canonical formalism. We discuss a few specific examples of gauges and make a correspondence between some popular choices of gauge made in the configuration space (the standard approach) and the gauge-fixing conditions put on the phase space (the Hamiltonian approach). This discussion is an important aspect of our formalism and it is not considered in \cite{Langlois:1994ec}. As our considerations are not motivated by the theory of inflation, we work with the universe filled with a cosmological fluid (or, many cosmological fluids). Our formalism can be used to quantize the dynamics of both the background and the perturbations in a consistent manner.

Since the Langlois paper there have been some useful developments in the Hamiltonian approach to cosmological perturbation theory. Let us mention some of them. For example, in \cite{Giesel:2017roz,Giesel:2018opa}, the ADM phase space is replaced with an extended phase space that encompasses the lapse, the shift and their conjugate momenta. The extended phase space formulation, though more complex, gives a more direct relation between gauge-fixing conditions in the phase space and in the usual (configuration space) approach. It also proves convenient for using the relational formalism \cite{Giesel:2008zz} to derive the gauge-invariant perturbations. 

The Hamiltonian approach to the universe with one or more perfect fluids was developed in \cite{Peter:2005hm, Pinho:2006ym, Peter:2015zaa} and a quantization of the entire system was proposed. Like our work, these very interesting results were driven by intent to study alternative to inflation scenarios of primordial universe. However, the authors obtained their results through working with the Lagrangian formalism and the Hamiltonian was only obtained in the last step, whereas our approach is based entirely on the canonical level.  Moreover, the problem of the relation between different choices of gauge was not much discussed in those works either.

For the Hamiltonian approach to second-order cosmological perturbation theory, see e.g. \cite{Domenech:2017ems}.

The outline of the present paper is as follows. In Sec. II we recall the Hamiltonian formalism for general relativity and relativistic fluids. In Sec. III we expand those formalisms in perturbations around the flat Friedmann-Lemaitre-Robertson-Walker (FLRW) universe and re-write the perturbations in the Fourier modes. In Sec. IV we discuss in detail the derivation of the reduced phase space equipped with the physical Hamiltonian by means of the Dirac procedure and we derive the physical Hamiltonians for vector and tensor modes of perturbations. In Sec. V we focus on scalar modes of perturbations. We consider a few examples of gauge-fixing conditions that correspond to some popular choices of gauge used in the configuration space approach. In Sec. VI we extend our result to the multi-fluid case. We conclude in Sec. VII. In Appendix \ref{expansion} the expansion of the ADM Hamiltonian up to second order is given. Appendix \ref{App1} describes the geometrical interpretation of the canonical variables used throughout the main text. Appendix \ref{App2} provides the relation between the variables of the present paper and the ones used in the configuration space approach to cosmological perturbation theory. {In particular, it provides the relation between the Dirac observables and the physical variables of the standard approach: the Bardeen potentials and the Mukhanov-Sasaki variable.}

\section{Canonical formalism}
In this section we briefly recall the basic elements of the ADM canonical formalism for the gravitational field and the canonical formalism for relativistic perfect fluids.
\subsection{Gravitational field}
In what follows we assume the topology of the spacetime $\mathcal{M}\simeq \mathbb{T}^3\times \mathbb{R}$ and the line element to read\footnote{ The purpose of assuming $\mathcal{M}\simeq \mathbb{T}^3\times \mathbb{R}$ is to have a spatially compact universe and to avoid the ambiguity in the definition of the symplectic structure for background (homogeneous) variables. We recall that for spatially non-compact universes one is forced to introduce an ambiguous finite fiducial cell whose choice influences the definition of the symplectic structure.}
\begin{equation}
ds^2=-N^2dt^2+q_{ab}(dx^a+N^adt)(dx^b+N^bdt),
\end{equation}
where $a,b=1,2,3$ are spatial coordinate indices. The Hilbert-Einstein action in the ADM variables reads
\begin{equation}
S_g=\int R\sqrt{-g}~d^4x=\int (^3R-K^2+K_a^{~b}K_b^{~a})N\sqrt{q}~\ud^3x\ud t,
\end{equation}
where $^3R$ is the Ricci scalar of the three-geometries (i.e. the level sets of $t$) and $K_{ab}$ is the extrinsic curvature tensor,
\begin{eqnarray}
K_{ab}=\frac{1}{2N}(\dot{q}_{ab}-D_aN_b-D_bN_a),
\end{eqnarray}
where $D_a$ is the spatial covariant derivative. The Legendre transformation $\dot{q}_{ab}\mapsto\pi^{ab}:=\frac{\delta S_g}{\delta \dot{q}_{ab}}$ establishes the ADM phase space,
\begin{align}
\left(q_{ab},\pi^{ab}=\sqrt{q}(K^{ab}-Kq^{ab})\right)\in\mathbb{R}_+^6\times\mathbb{R}^6,~\{q_{ab}(x),\pi^{cd}(x')\}=\delta_{(a}^{~(c}\delta_{b)}^{~d)}\delta(x-x'),
\end{align}
in which the dynamics is generated by the Hamiltonian,
\begin{align}
{\bf H}_g=\int (N \mathcal{H}_{g,0} +N^a\mathcal{H}_{g,a})~\ud^3x,
\end{align}
which is a sum of first-class constraints,
\begin{eqnarray}
\mathcal{H}_{g,0}=\sqrt{q}\left(-^3R+q^{-1}(\pi_a^{~b}\pi_b^{~a}-\frac{1}{2}\pi^2)\right),~~
\mathcal{H}^{~b}_g=-2D_a(\pi^{ab}),
\end{eqnarray}
where the lapse $N$ and the shifts $N^a$ play the role of the Lagrange multipliers. 

\subsection{Relativistic fluids}
We extend the phase space in order to include perfect fluids, each satisfying the barotropic equation of state $p=w\rho$, where  $p$ is pressure, $\rho$ is energy density in the frame comoving with the fluid and $w$ is a constant. Inspired by the Schutz velocity-potential approach to the variational formulation of relativistic perfect fluids \cite{Schutz:1970my,Schutz:1971ac,Demaret:1980nq} we start from the action,
\begin{equation}\label{actionF0}
S_f=\int \sqrt{g}p(\mu,S)\ud^4 x=\int N \sqrt{q}p(\mu,S)\ud^3 x\ud t=:\int N\mathcal{L}_f\ud^3 x\ud t,
\end{equation}
where $\mu=\frac{\rho+p}{n}$ is enthalpy per fluid's particle (it plays the role of the inertial mass of a fluid's particle), $n$ is the number density of fluid's particles and $S$ is entropy per fluid's particle. The first law of thermodynamics expressed in terms of $p$, $S$, $\mu$ and $n$ reads,
 \begin{align}\label{1stlt}
 \ud p=n\ud\mu-nT\ud S,
 \end{align} 
 where $T$ is the fluid temperature. As perfect fluids do not conduct heat, the entropy of each element of the fluid is conserved. Therefore, we will assume the specific entropy $S$ to be homogeneous across the space and hence to remain constant in time despite the flow of the fluid. Furthermore, we will assume rotationless motion of the fluid's particles. Then, the four-velocity of the fluid can be assumed to be a function of the following form,
 \begin{equation}\label{schutz0}
U^{\nu}=\mu^{-1}\phi_{,}^{~\nu},~~U_{\nu}U^{\nu}=-1
\end{equation}
i.e. to be a function of a single scalar field $\phi$ and the four-metric. Notice that the normalization condition for $U^{\nu}$ determines the specific enthalpy $\mu$ in terms of $\phi$, $q_{ab}$, $N$ and $N^a$. Minimization of the action (\ref{actionF0}) with respect to the metric leads to
\begin{align}
\frac{\delta S_f}{\delta g_{\mu\nu}}=\frac{\sqrt{g}}{2}\left(pg^{\mu\nu}+(\rho+p)U^{\mu}U^{\nu}\right)=\frac{\sqrt{g}}{2}T^{\mu\nu},
\end{align} 
i.e. to the stress-energy tensor of a perfect fluid, whereas minimization with respect to the scalar field $\phi$ leads to
\begin{align}
\frac{\delta S_f}{\delta \phi}=\sqrt{g}(nU^{\mu})_{;\mu}=0,
\end{align}
i.e. to the fluid flow that conserves the number of particles. Combined with the conservation of the stress-energy tensor $T^{\mu\nu}_{~~;\nu}=0$, the above equation determines the motion of the perfect fluid whose equation of state is known. 

The Legendre transformation, $\dot{\phi}\mapsto p^{\phi}:=\frac{\delta S_f}{\delta\dot{\phi}}$, yields the phase space,
\begin{equation}
\left(\phi, p^{\phi}=-N\sqrt{q}nU^0\right)\in\mathbb{R}_+\times\mathbb{R},~\{\phi(x),p^{\phi}(x')\}=\delta(x-x'),
\end{equation}
where  we used the definition of $\mu$ (\ref{schutz0}) and the relation of the number density to the pressure, $n=\frac{\partial p}{\partial \mu}\big|_{S=const}$, via the first law of thermodynamics (\ref{1stlt}). The Hamiltonian reads:
\begin{equation}
{\bf H}_f=\int(\dot{\phi}p^{\phi}-N \mathcal{L}_f)~\ud^3x=\int N\sqrt{q}(-(\rho+p)U^0U_0-p)~\ud^3x=-\int N\sqrt{q}T^0_{~0}.
\end{equation}
We will assume the equation of state,
\begin{equation}\label{pres}
p(\mu)=K\mu^{\alpha},~~\alpha=\frac{w+1}{w},
\end{equation}
where $K$ is an arbitrary constant and which is equivalent to $p=w\rho$. This specified equation of state specifies the definition of momentum $p^{\phi}$ as follows:
\begin{equation}
\frac{1}{N}\phi_{,0}=\frac{p^{\phi}}{\sqrt{q}K\alpha\mu^{\alpha-2}}+\frac{N^a}{N}\phi_{,a}~,
\end{equation}
where the specific enthalpy $\mu$ can be determined as a function of the momentum $p^{\phi}$, the field ${\phi}$ and the spatial metric $q_{ab}$ by making use of the normalization condition for $U^{\mu}$:
\begin{equation}\label{mu2}
-\mu^2=-\left(\frac{p^{\phi}}{\sqrt{q}K\alpha\mu^{\alpha-2}}\right)^2+q^{ab}\phi_{,a}\phi_{,b}.
\end{equation}
Although, finding the explicit solution to this equation might be difficult, we do not really need it and we will keep $\mu$ in the sequel. The Hamiltonian is found to read:
\begin{equation}
{\bf H}_f=\int(N\frac{(p^{\phi})^2}{\sqrt{q}K\alpha\mu^{\alpha-2}}-N\sqrt{q}K\mu^{\alpha}+N^ap^{\phi}\phi_{,a})~\ud^3x~.
\end{equation}
Hence, the fluid constraints read:
\begin{equation}
\mathcal{H}_{f,0}=\frac{(p^{\phi})^2}{\sqrt{q}K\alpha\mu^{\alpha-2}}-\sqrt{q}K\mu^{\alpha},~~\mathcal{H}_{f,a}=p^{\phi}\phi_{,a}.
\end{equation}
For the physical interpretation of the fluid part of the Hamiltonian let us introduce a new basis for space-time vector and co-vector fields, 
\begin{align}\label{vecbasis}
e_{\bar{0}}:=\frac{1}{N}\partial_{t}-\frac{N^a}{N}\partial_{x^a},~~e_{\bar{a}}:=\partial_{x^a},~~e^{\bar{0}}=N\ud t,~~e^{\bar{a}}=N^a\ud t+\ud x^a,~~e^{\bar{\mu}}(e_{\bar{\nu}})=\delta^{\bar{\mu}}_{~\bar{\nu}},
\end{align}
where we denote the new basis with the barred indices, $\bar{0}$ and $\bar{a}$, to distinguish it from the coordinate basis denoted with the unbarred indices, ${0}$ and $a$. Note that $e_{\bar{0}}$ is the normal vector to the constant-time surface and it is normalized. The vectors $e_{\bar{a}}$ are tangent to that surface and thus, orthogonal to $e_{\bar{0}}$. They are not normalized as $e_{\bar{a}}\cdot e_{\bar{b}}=q_{ab}$. We find that
\begin{align}
\mathcal{H}_{f,0}=\sqrt{q}\left(-p-(\rho+p)U^{\bar{0}}U_{\bar{0}}\right)=-\sqrt{q} T^{\bar{0}}_{~\bar{0}},~~\mathcal{H}_{f,a}=-\sqrt{q}(\rho+p)U^{\bar{0}}U_{\bar{a}}=-\sqrt{q} T^{\bar{0}}_{~~\bar{a}},
\end{align}
where $T^{\bar{0}\bar{0}}=-T^{\bar{0}}_{~\bar{0}}$ is the energy density of the fluid relative to the normal, $e_{\bar{0}}$ (it is not the same with the energy density in the fluid's frame), and $T^{\bar{0}}_{~~\bar{a}}$ is the flow of energy through the surfaces of the constancy of $ e^{\bar{a}}$ (and measured relatively to $e_{\bar{0}}$).
Finally, let us notice that the fluid four-vector $U^{\mu}$ expressed in terms of the canonical variables reads,
\begin{align}
U^0=-\frac{p^{\phi}}{N\sqrt{q}K\alpha\mu^{\alpha-1}},~~U^a=\frac{p^{\phi}}{N\sqrt{q}K\alpha\mu^{\alpha-1}}N^a+\mu^{-1}\phi_{,b}q^{ba},
\end{align}
and hence, the fluid flow is orthogonal to constant time surfaces if and only if
\begin{align}\label{orthogonal}
\vec{U}\cdot e_{\bar{a}}=\mu^{-1}\phi_{,a}=0,~~\textrm{for}~~a=1,2,3.
\end{align}
The total gravity + fluid Hamiltonian reads:
\begin{align}
{\bf H}={\bf H}_g+{\bf H}_f=\int (N \mathcal{H}_{g,0}+N \mathcal{H}_{f,0} +N^a\mathcal{H}_{g,a}+N^a\mathcal{H}_{f,a})~\ud^3x.
\end{align}
\section{Perturbative expansion of canonical formalism}
In this section we expand the above canonical formalism in perturbations to the flat FLRW universe. The canonical background variables read
\begin{equation}\label{defa0}
a^2=\frac{1}{3v_0^{1/3}}\int q_{ab}\delta^{ab}~\ud^3x,~~~p=\frac{1}{v_0^{2/3}}\int\pi^{ab}\delta_{ab}~\ud^3x,~~~\bar{\phi}=\frac{1}{v_0^{1/3}}\int\phi~\ud^3x,~~~\bar{p}^{\phi}=\frac{1}{v_0^{2/3}}\int p^{\phi}~\ud^3x,
\end{equation}
where $v_0:=\int\ud^3x$ is the coordinate volume of the spatial leaf $\mathbb{T}^3$. Let us assume $v_0=1$. The canonical perturbation variables read
\begin{align}\label{defP0}
\delta q_{ab}=q_{ab}-a^2\delta_{ab}
,~~\delta\pi^{ab}=\pi^{ab}-\frac{1}{3}p\delta^{ab},~~\delta\phi=\phi-\bar{\phi},~~\delta p^{\phi}=p^{\phi}-\bar{p}^{\phi},
\end{align}
and the Poisson brackets read (with all the remaining commutation relations vanishing)
\begin{align}\label{PB1}\begin{split}
\{\delta \phi(x),\delta p^{\phi}(x')\}=\delta^3(x-x'),~~\{\delta q_{ab}(x),\delta\pi^{cd}(x')\}=\delta_{(a}^{~c}\delta_{b)}^{~d}\delta^3(x-x'),\\
\{\bar{\phi},\bar{p}^{\phi}\}=1,~~\{a^2,p\}=1.\end{split}
\end{align}
We also introduce the perturbations of the lapse and the shifts that are not any part of the phase space via the replacement: $N\mapsto N+\delta N$ and $N^a\mapsto N^a+\delta N^a$, respectively, where $N$ and $N^a$ are now understood as zero-order quantities. 

The total Hamiltonian expanded up to second order reads
\begin{equation}\label{hamtot}
{\bf H}=N\mathcal{H}_{0}\big|^{(0)}+\int (N\mathcal{H}_{0}\big|^{(2)}+\delta N\mathcal{H}_{0}\big|^{(1)}+\delta N^a\mathcal{H}_{a}\big|^{(1)})~\ud^3x,
\end{equation}
where the terms $\mathcal{H}_{0}\big|^{(0)}$, $\mathcal{H}_{0}\big|^{(2)}$, $\mathcal{H}_{0}\big|^{(1)}$ and $\mathcal{H}_{a}\big|^{(1)}$ of the above expansion are given in Appendix \ref{expansion}. Notice that strictly speaking, the above Hamiltonian does not define a gauge system as the first-order constraints $\mathcal{H}_{0}\big|^{(1)}$ and $\mathcal{H}_{a}\big|^{(1)}$'s do not commute beyond first order. Only the extra assumption that the dynamics which it generates is to be truncated at linear order makes the dynamics of perturbations subject to gauge transformations.

We Fourier-transform the perturbation variables (see e.g. \cite{Dapor:2013pka}),
\begin{equation}
\delta\check{q}_{ab}(\vec{k})=\int \delta {q}_{ab}(\vec{x})e^{-i\vec{x}\vec{k}}d^3x, ~~~~~~\delta\check{\pi}^{ab}(\vec{k})=\int \delta{\pi}^{ab}(\vec{x})e^{-i\vec{x}\vec{k}}d^3x,
\end{equation} 
where the reality conditions $\delta\check{q}_{ab}(-\vec{k})=\overline{\delta\check{q}}_{ab}(\vec{k})$ and $\delta\check{\pi}^{ab}(-\vec{k})=\overline{\delta\check{\pi}}^{ab}(\vec{k})$ are satisfied. Furthermore, we express the tensors in a basis that splits the modes into scalars, vectors and tensors. First, let us introduce two vectors, $\vec{v}$ and $\vec{w}$, orthogonal to each other and to $\vec{k}$, whose magnitude in the fiducial metric $\delta_{ab}$ reads,
\begin{align}
|\vec{v}|=|\vec{w}|=|\vec{k}|^{-1}.
\end{align} 
Also, we assume that our definition of the frame is symmetric with respect to the reflection about the origin, that is, when $\vec{k}\rightarrow -\vec{k}$ then $\vec{v}\rightarrow -\vec{v}$ and $\vec{w}\rightarrow -\vec{w}$. Next, we define
\begin{equation}
\delta\check{q}_{ab}=\delta\check{q}_{m}A_{ab}^m,~~~~~~\delta\check{\pi}^{ab}=\delta\check{\pi}^{m}A_{m}^{ab},
\end{equation}
where $A_{ab}^1:=\delta_{ab}$, $A_{ab}^2:=\frac{k_a k_b}{k^2}-\frac{1}{3}\delta_{ab}$, $A_{ab}^3:=\frac{1}{\sqrt{2}}(k_a v_b+k_b v_a)$, $A_{ab}^4:=\frac{1}{\sqrt{2}}(k_a w_b+k_b w_a)$, $A_{ab}^5:=\frac{k^2}{\sqrt{2}}(v_a w_b+v_b w_a)$, $A_{ab}^6:=\frac{k^2}{\sqrt{2}}(v_a v_b-w_a w_b)$. The matrices $A^{ab}_m$ form the dual basis, i.e. $A^{ab}_mA_{ab}^n=\delta^n_m$. The new variables describe respectively scalar ($\delta\check{q}_{1}$ and $\delta\check{q}_{2}$), vector ($\delta\check{q}_{3}$ and $\delta\check{q}_{4}$) and tensor ($\delta\check{q}_{5}$ and $\delta\check{q}_{6}$) modes of the metric perturbation. The Poisson brackets (\ref{PB1}) take the form
\begin{align}\label{PB}
\{ \delta q_i(\vec{k}),\delta\pi^j(-\vec{l})\}=\delta_i^{~j}\delta_{\vec{k},\vec{l}},~~\{ \delta \phi(\vec{k}),\delta p^{\phi}(-\vec{l})\}=\delta_{\vec{k},\vec{l}},
\end{align}
where $i,j=1,\dots,6$, and where we omit the zero-order commutation relations (from now on we omit the inverted hat over the Fourier modes of perturbations). 

The zero-order terms in the Hamiltonian (\ref{hamtot}) read,
\begin{align}
\mathcal{H}_{0}\big|^{(0)}=-\frac{1}{6}ap^2+(\alpha-1)a^3K\left(\frac{|\bar{p}^{\phi}|}{K\alpha a^3}\right)^{\frac{\alpha}{\alpha-1}},
\end{align}
while the Fourier transformation of the first- and second-order terms in (\ref{hamtot}) yields,
\begin{align}
\mathcal{H}_{g,0}\big|^{(1)}&=-\frac{ap}{3}\delta\pi^1-a^{-1}\left(\frac{p^2}{12}+2k^2\right)\delta q_1+\frac{2}{3}a^{-1}k^2\delta q_2,
\end{align}
\begin{align}
\mathcal{H}_{g,a}\big|^{(1)}&=ik_aa^2\left(-\frac{2}{3}\delta\pi^1-2\delta\pi^2+\frac{a^{-2}p}{3}(\delta q_1-\frac{4}{3}\delta q_2)\right),
\end{align}
\begin{align}\begin{split}
\mathcal{H}_{g,0}\big|^{(2)}&=-\frac{1}{6}a(\delta\pi^1)^2+\frac{3}{2}a(\delta\pi^2)^2-\frac{1}{6}a^{-1}p\delta\pi^1\delta q_1+\frac{1}{3}a^{-1}p\delta\pi^2\delta q_2+\frac{1}{48}a^{-3}p^2(\delta q_1)^2\\  &+\frac{5}{108}a^{-3}p^2(\delta q_2)^2-\frac{1}{2}a^{-3}k^2(\delta q_1)^2-\frac{1}{18}a^{-3}k^2(\delta q_2)^2+\frac{1}{3}a^{-3}k^2(\delta q_1)(\delta q_2)\\   &+a[(\delta\pi^3)^2+(\delta\pi^4)^2]+\frac{a^{-1}p}{3}[(\delta\pi^3\delta q_3)+(\delta\pi^4\delta q_4)]\\ &+\left(\frac{5+3w}{72}a^{-3}p^2+\frac{a^{-3}k^2}{2}\right)[(\delta q_3)^2+(\delta q_4)^2]\\ 
&+ a[(\delta\pi^5)^2+(\delta\pi^6)^2]+\frac{a^{-1}p}{3}[(\delta\pi^5\delta q_5)+(\delta\pi^6\delta q_6)]\\ &+\left(\frac{5+3w}{72}a^{-3}p^2+\frac{a^{-3}k^2}{4}\right)[(\delta q_5)^2+(\delta q_6)^2],\end{split}
\end{align}
\begin{align}
\mathcal{H}_{f,0}\big|^{(1)}&=K\alpha a^3\left(\frac{|\bar{p}^{\phi}|}{K\alpha a^3}\right)^{\frac{\alpha}{\alpha-1}}\left(\frac{\delta p^{\phi}}{\bar{p}^{\phi}}-\frac{3}{2\alpha}\frac{\delta q_1}{a^2}\right),
\end{align}
\begin{align}
\mathcal{H}_{f,a}\big|^{(1)}&=ik_a\bar{p}^{\phi}\delta\phi~,
\end{align}
\begin{align}\begin{split}
&\mathcal{H}_{f,0}\big|^{(2)}=K\sqrt{q}\mu^{\alpha}\alpha\\
\times&\left(\frac{1}{2(\alpha-1)}\left(\frac{\delta p^{\phi}}{\bar{p}^{\phi}}\right)^2-\frac{3}{2(\alpha-1)}\left(\frac{\delta q_1}{a^2}\right)\left(\frac{\delta p^{\phi}}{\bar{p}^{\phi}}\right)+\frac{9}{8\alpha(\alpha-1)}\left(\frac{\delta q_1}{a^2}\right)^2+\frac{1}{4\alpha}\frac{3(\delta q_1)^2+\frac{2}{3}(\delta q_2)^2}{a^4}\right)\\
&~~~~~~~+k^2K\sqrt{q}\mu^{\alpha-2}\frac{\alpha}{2}a^{-2}\left(\delta\phi\right)^2,\end{split}
\end{align}
where $\mu^{\alpha}=\left(\frac{|\bar{p}^{\phi}|}{K\alpha a^3}\right)^{\frac{\alpha}{\alpha-1}}$ is zero-order. Notice that the above second-order terms are the products of $(+\vec{k})$- and $(-\vec{k})$-modes. 

\section{Dynamics and gauge-fixing conditions}\label{IV}
The dynamics of the entire system, i.e. of both the background and the perturbations, is generated by the total Hamiltonian (\ref{hamtot}). It includes one zero-order and four first-order constraints, namely\footnote{ Following Dirac’s nomenclature we say that two phase space functions are weakly (strongly) equal when they are equal in the constraint surface (all over the phase space).}
\begin{align}
\mathcal{H}_{0}\big|^{(0)}\approx 0,~~\mathcal{H}_{\mu}\big|^{(1)}\approx 0,~\mu=0,1,2,3.
\end{align}
At zeroth order the Hamiltonian is a constraint, which gives rise to the so-called time problem. Namely, the naive application of the Dirac procedure results in physical observables which are constants of motion, i.e. they commute with the constraint, and hence the dynamics appears to be a pure gauge transformation. The most common approaches to the time problem are discussed in Isham \cite{Isham:1992ms} and Kuchar \cite{Kuchar:1991qf}. The time problem, however, is not the subject of the present paper and interested readers are referred to our previous papers \cite{0264-9381-32-13-135004,Malkiewicz:2015fqa,0264-9381-34-14-145012,PhysRevD.96.046003}. In what follows we focus on the second-order part of the Hamiltonian. Let us rewrite the first-order constraints in the basis $(\vec{k},\vec{v},\vec{w})$. The diffeomorphism constraints read
\begin{align}\nonumber
\mathcal{H}_{\vec{k}}\big|^{(1)}&:=\mathcal{H}_{a}\big|^{(1)}k^ak^{-2}=-i2a^2\left(\frac{1}{3}\delta\pi_1+\delta\pi_2+\frac{pa^{-2}}{3}\left(\frac{2}{3}\delta q_2-\frac{1}{2}\delta q_1\right)\right)+i\bar{p}^{\phi}\delta\phi,\\ \label{diffeo}
\mathcal{H}_{\vec{v}}\big|^{(1)}&:=\mathcal{H}_{a}\big|^{(1)}v^a=-\frac{i2a^2}{\sqrt{2}}\left(\delta\pi_3+\frac{pa^{-2}}{3}\delta q_3\right),\\ \nonumber
\mathcal{H}_{\vec{w}}\big|^{(1)}&:=\mathcal{H}_{a}\big|^{(1)}w^a=-\frac{i2a^2}{\sqrt{2}}\left(\delta\pi_4+\frac{pa^{-2}}{3}\delta q_4\right),
\end{align}
whereas the scalar constraint reads
\begin{align}\label{scalar}
\mathcal{H}_{0}\big|^{(1)}=-\frac{ap}{3}\delta\pi_1-\frac{a^{-1}p^2}{12}\delta q_1+a^{-1}k^2\left(\frac{2}{3}\delta q_2-2 \delta q_{1}\right)+K\alpha a^3\left(\frac{|\bar{p}^{\phi}|}{K\alpha a^3}\right)^{\frac{\alpha}{\alpha-1}}\left(\frac{\delta p^{\phi}}{\bar{p}^{\phi}}-\frac{3}{2\alpha}\frac{\delta q_1}{a^2}\right).
\end{align}
Notice that $\mathcal{H}_{\vec{k}}\big|^{(1)}$ is the scalar mode of the diffeomorphism constraint, whereas $\mathcal{H}_{\vec{v}}\big|^{(1)}$ and $\mathcal{H}_{\vec{w}}\big|^{(1)}$ are the vector modes. The scalar constraint is made of scalar modes only. The lapse function $\delta N$ is a purely scalar quantity, whereas the shift vector may be expressed in the basis $(\vec{k},\vec{v},\vec{w})$ as  $\delta \vec{N}=\delta N^{\vec{k}}+\delta N^{\vec{v}}+\delta N^{\vec{w}}$, where the longitudinal part of the shift vector, $\delta N^{\vec{k}}=i\delta n\vec{k}$, is a scalar mode and $\delta N^{\vec{v}}$, $\delta N^{\vec{w}}$ are vector modes. The total Hamiltonian (\ref{hamtot}) splits into scalar, vector and tensor parts, each generating dynamics exclusively of the respective modes,
\begin{equation}
{\bf H}^{(2)}=\int (N\mathcal{H}_{0}\big|^{(2S)}+N\mathcal{H}_{0}\big|^{(2V)}+N\mathcal{H}_{0}\big|^{(2T)}+\delta N\mathcal{H}_{0}\big|^{(1)}-k^2\delta n\mathcal{H}_{\vec{k}}\big|^{(1)}+\delta N^{\vec{v}}\mathcal{H}_{\vec{v}}\big|^{(1)}+\delta N^{\vec{w}}\mathcal{H}_{\vec{w}}\big|^{(1)}),
\end{equation}
where $\mathcal{H}_{0}\big|^{(2S)}$, $\mathcal{H}_{0}\big|^{(2V)}$ and $\mathcal{H}_{0}\big|^{(2T)}$ denote respectively the scalar, the vector and the tensor part of the nonvanishing second-order term $\mathcal{H}_{0}\big|^{(2)}$.

\subsection{Gauge-fixing conditions and Dirac brackets}
The above constraints are first-class in the sense that they weakly commute with each other up to first order\footnote{Actually, they commute strongly up to first order.}. Therefore, in the space of linear perturbations, these constraints (a) define the physical surface in the phase space and (b) generate gauge transformations via the respective vector fields, denoted by 
\begin{align}\label{GT}
\delta\vec{\xi}(\cdot)=\{\cdot~,\int\delta{\xi}^{\mu}\mathcal{H}_{\mu}\big|^{(1)}\},~~\mu=0,\vec{k},\vec{v},\vec{w}
\end{align}  
where $\delta{\xi}^{\mu}$ are first order. Those vector fields commute with each other in the physical surface\footnote{It follows from the fact that the respective constraints commute weakly and that there exists a homomorphism between the phase space observables and their Hamiltonian vector fields by the virtue of the Jacobi identity.}. By the virtue of the Frobenius theorem they generate sub-manifolds in the constraint surface that are called gauge orbits. The Dirac procedure starts with choosing gauge-fixing conditions, denoted by
\begin{align}
\delta c_{\mu}\approx 0,~~~\mu=0,\vec{k},\vec{v},\vec{w}.
\end{align} 
They are such that the elements of the total set of the constraints, denoted by 
\begin{align}
\phi_{\mu}=\mathcal{H}_{0}\big|^{(1)},\dots,\mathcal{H}_{\vec{w}}\big|^{(1)},\delta c_{0},\dots,\delta c_{\vec{w}},
\end{align} 
form an invertible matrix of the commutation relations,
\begin{align}\label{invB}
\textrm{det} |\{\phi_{\mu},\phi_{\nu}\}|\neq 0.
\end{align}
If the above condition is satisfied, then one introduces the Dirac brackets to replace the Poisson brackets with,
\begin{align}\label{db1}
\{\cdot,\cdot\}_D=\{\cdot,\cdot\} -\{\cdot,\phi_{\mu}\}\{\phi_{\mu},\phi_{\nu}\}^{-1}\{\phi_{\nu},\cdot\}.
\end{align}
One can easily verify that $\{\phi_{\mu},\cdot\}_D=0$ for any $\phi_{\mu}$. In other words, the Dirac brackets are blind to the constraints and gauge-fixing conditions. Therefore, one can replace the weak equalities $\phi_{\mu}\approx 0$ with the strong ones $\phi_{\mu} = 0$. More specifically, one can use the constraints to reduce the number of canonical variables {\it before} evaluating the Dirac brackets. Such a reduced set of variables parametrize the gauge-fixing surface whose canonical structure is given by the Dirac brackets. The physical Hamiltonian in the gauge-fixing surface is simply the initial ADM Hamiltonian (\ref{hamtot}) to which the strong constraints $\phi_{\mu} = 0$ are applied. We notice that the values of $\delta N$ and $\delta N^a$ become irrelevant in Eq. (\ref{hamtot}) as $\mathcal{H}_{\mu}\big|^{(1)}$'s vanish in the gauge-fixing surface which is contained in the constraint surface. However, for the full reconstruction of the metric one needs to know their values. This can be achieved by noticing that the gauge-fixing conditions must be consistent with the dynamics, i.e.
\begin{equation}\label{stabilityEq}
\{\delta c_{\mu},{\bf H}\} \approx 0,~~~\mu=0,\vec{k},\vec{v},\vec{w}.
\end{equation}
(Notice that now we use the Poisson brackets). We will refer to the above equation as the consistency condition. This is an algebraic equation for $\delta N$ and $\delta N^a$ as will be seen shortly.

\subsection{Canonical isomorphism}
\begin{figure}[t]
\includegraphics[width=0.8\textwidth]{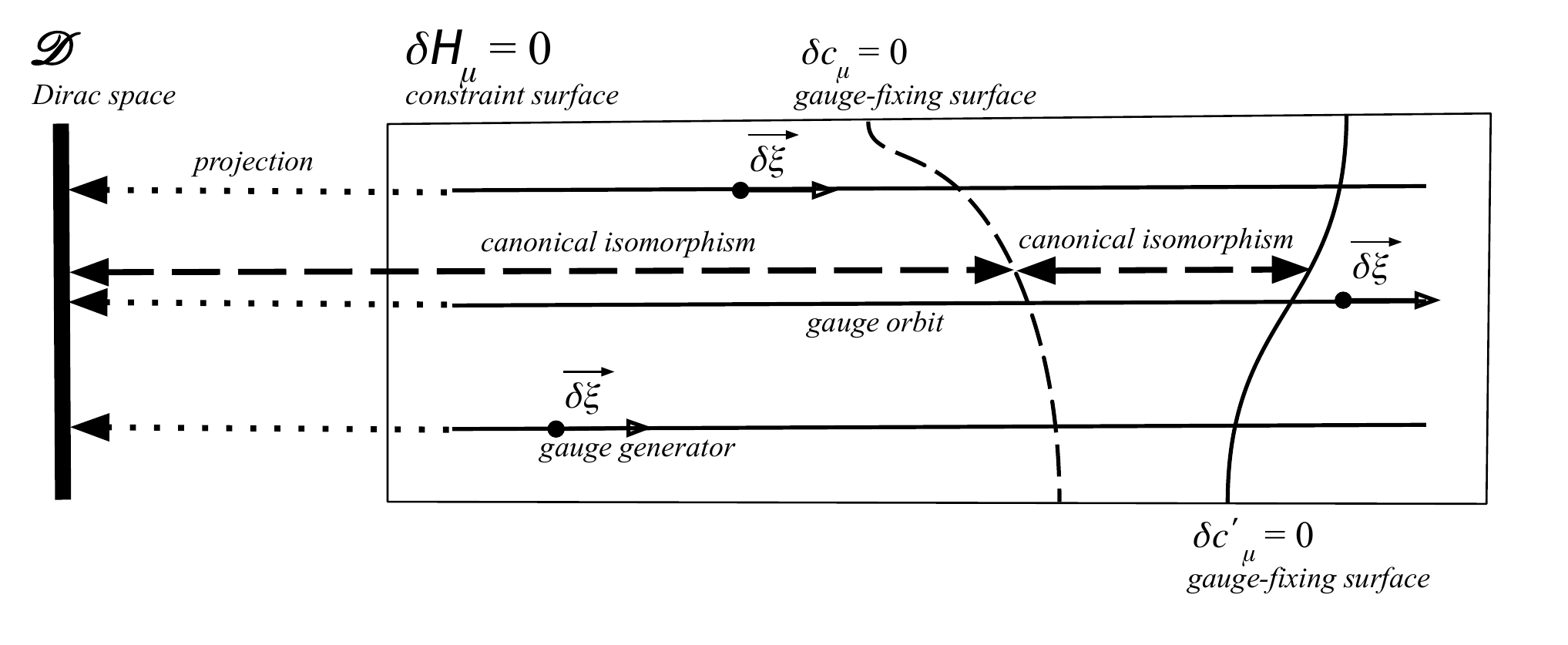}
\caption{ Geometrical aspects of the Dirac gauge-fixing procedure for constrained systems: The first-class constraints induce the constraint surface in the kinematical phase space, $\delta\mathcal{H}_{\mu}= 0$. Their Hamiltonian vector fields $\delta{\xi}^{\mu}$ generate sub-manifolds in the constraint surface, which are called the gauge orbits. Any two points in a given gauge orbit represent the same physical state of the constrained system. A complete set of gauge-fixing conditions, $\delta c_{\mu}= 0$, defines a smooth gauge-fixing surface in the constraint surface that contains one point per each gauge orbit. The physical motion is confined to the gauge-fixing surface by means of the Dirac brackets that are blind to the constraints and the gauge-fixing conditions. The same physical motion can be also considered in the quotient space of all gauge orbits, the Dirac space $\mathcal{D}$. Coordinates on the quotient space are identical with Dirac observables, that is, observables that are constant along each gauge orbit. There exists a canonical isomorphism between the quotient space $\mathcal{D}$ equipped with the commutation relations between the Dirac observables and any gauge-fixing surface $\delta c_{\mu}= 0$ equipped with the respective Dirac brackets.}
\label{fig1}
\end{figure}

The next issue we should address is the relation between different choices of the gauge-fixing conditions, or different gauge-fixing surfaces. In what follows, we show that every gauge-fixing surface represents the {\it universal} physical phase space. Let us observe that there must exist the so called Dirac observables (assumed of first order), denoted by $\delta D_i\in\mathcal{D}$, that weakly commute with the first-class constraints,
\begin{align}\label{defdir}
\forall_{\delta{\xi}^{\mu}}~\{\delta D_i~,\int\delta{\xi}^{\mu}\mathcal{H}_{\mu}\big|^{(1)}\}\approx 0,
\end{align} 
and are not the first-class constraints themselves. Since the above equality is weak, one can think of Dirac observables either as equivalence classes of functions on the kinematical phase space, which satisfy Eq. (\ref{defdir}) and coincide on the constraint surface, or as particular representatives of the equivalence classes. It is straightforward to show that (a) they form a closed algebra in the constraint surface (weakly)\footnote{To show this it is sufficient to invoke the Jacobi identity.}, i.e. given $\delta D_i$ and $\delta D_j$, $\{\delta D_i~,\delta D_j\}\approx \delta D_k$ for some $\delta D_k$ and (b) that this algebra can be computed in {\it any} Dirac brackets (i.e., based on {\it any} choice of gauge-fixing conditions) as well as in the Poisson brackets, i.e. 
\begin{align}\label{Dalgebra}
\{\delta D_i~,\delta D_j\}_D\approx\{\delta D_i~,\delta D_j\}.
\end{align}
On the one hand, it is clear that the number of Dirac observables must be equal to the number of the reduced variables parametrizing any gauge-fixing surface. On the other hand, any variable in the gauge-fixing surface must be equal to a certain Dirac observable modulo the constraints and gauge-fixing conditions that are however invisible to the Dirac brackets (\ref{db1}). Therefore, we conclude that there must exist a canonical isomorphism between the reduced set of variables in any gauge-fixing surface and the set of Dirac observables (see below for more details). Suppose that for a given set of gauge-fixing conditions, $\delta c_{\mu}= 0$, $\mu=0,\vec{k},\vec{v},\vec{w}$, one has chosen a reduced set of canonical variables, denoted by $\{v_i\}$. Then the isomoprhism is given by the invertible map
\begin{align}
\mathcal{D}\ni\delta D_i\mapsto \mathcal{O}_i(v_j)= \delta D_i\bigg|_{\delta c_{\mu}= 0}~,
\end{align}
where the equality between the Dirac observables $\delta D_i$ and the respective phase space observables $\mathcal{O}_i$ holds in a given gauge-fixing surface only. Since this map is a canonical isomorphism it can be used to map the physical Hamiltonian to the space of Dirac observables. Therefore, we conclude that there exists a unique physical phase space, parametrized by the Dirac observables, with the dynamics generated by a unique physical Hamiltonian that is a function of Dirac observables. The choice of gauge-fixing conditions merely gives a representation to the Dirac observables in terms of a reduced set of the canonical variables from the kinematical phase space. Also, notice that the fundamental canonical isomorphism can be used to establish a canonical relation between any two gauge-fixing surfaces. The Dirac procedure is depicted in Fig. \ref{fig1}.

Finally, in order to have perhaps a still better understanding of the Dirac brackets let us notice that they can be equivalently defined as
\begin{align}\label{altdir}
\{\cdot,\cdot\}_D=\{\cdot,\delta D_i\}\{\delta D_i,\delta D_j\}^{-1}\{\delta D_j,\cdot\},
\end{align}
where $\delta D_i$'s form a complete set of independent observables such that
\begin{align}\label{diracobs}
\{\delta D_i,\phi_{\mu}\}=0,~~~\textrm{for all}~\mu.
\end{align} 
In other words, $\delta D_i$'s are such Dirac observables that commute with the given gauge-fixing conditions, $\delta c_{\mu}=0$. Observe that the notion of Dirac observables applies only to the constraint surface functions and that the condition that Dirac observables must commute with the given gauge-fixing conditions does not reduce the number of Dirac observables. Instead, it specifies the way the Dirac observables as functions of the kinematical phase space extend beyond the constraint surface, that is, the mentioned condition determines a particular representation of Dirac observables as kinematical phase space functions. One can easily verify that given {\it any} set of Dirac observables, say $\delta\bar{D}_i$, they can be turned into $\delta D_i=\delta\bar{D}_i+\lambda^{\mu}\mathcal{H}_{\mu}\big|^{(1)}$ which satisfy Eq. (\ref{diracobs}) in the neighbourhood of the constraint surface, since the equations
\begin{align}
\{\delta\bar{D}_i+\lambda^{\mu}\mathcal{H}_{\mu}\big|^{(1)},\delta c_{\nu}\}\approx \{\delta\bar{D}_i,\delta c_{\nu}\}+\lambda^{\mu}\{\mathcal{H}_{\mu}\big|^{(1)},\delta c_{\nu}\}=0, ~~\textrm{for all}~\nu,
\end{align} 
have a unique solution $\lambda^{\mu}$ as $\{\mathcal{H}_{\mu}\big|^{(1)},\delta c_{\nu}\}$ is an invertible matrix. If one wants to use the Dirac observables to define a new coordinate system on the kinematical phase space, as did Langlois in \cite{Langlois:1994ec}, one needs to specify how the Dirac observables extend beyond the constraint surface and this is equivalent to the choice of gauge-fixing conditions. This explains the origin of and the way to deal with the ambiguity of the canonical transformation in the Langlois procedure. The fact that Langlois obtained a simple physical Hamiltonian which did not contain the curvature term is now seen as an implication of his choice of Dirac observables that commute with such gauge-fixing conditions that kill the curvature term. Such a set of gauge-fixing conditions is known as the spatially flat slicing (or, uniform curvature) gauge. We will discuss it in the next section.

The formula (\ref{altdir}) enables to understand how the Dirac brackets operate on observables in the kinematical phase space. Namely, any observable put into the brackets is interpreted as a Dirac observable (\ref{diracobs}) that coincides with the given observable in the gauge-fixing surface. Then, the result of the commutation is the respective Dirac observable in accordance with the universal commutation rules given in Eq. (\ref{Dalgebra}) and represented again by the respective function of the form (\ref{diracobs}).

\subsection{Dynamics of vector modes}
Let us now consider two of the diffeomorphism constraints (\ref{diffeo}), namely $\mathcal{H}_{\vec{v}}\big|^{(1)}\approx 0$ and $\mathcal{H}_{\vec{w}}\big|^{(1)}\approx 0$. Their vanishing imposes relations between the canonical variables of each pair $(\delta q_3,\delta \pi_3)$ and $(\delta q_4,\delta \pi_4)$, namely
\begin{align}\label{vecH}
\delta \pi_3+\frac{pa^{-2}}{3}\delta q_3\approx 0,~~\delta \pi_4+\frac{pa^{-2}}{3}\delta q_4\approx 0.
\end{align}
The respective gauge transformations (\ref{GT}) generated by these constraints can be fixed with the following conditions
\begin{align}\label{vecG}
\delta c_{\vec{v}}:=\delta q_3\approx 0,~~\delta c_{\vec{w}}:=\delta q_4\approx 0.
\end{align}
One may immediately verify that when taken together they all become second-class and that Eq. (\ref{invB}) indeed holds. One easily evaluates $\{\cdot,\phi_{A}\}\{\phi_{A},\phi_{A'}\}^{-1}\{\phi_{A'},\cdot\}$, $A\in\{\vec{v},\vec{w}\}$, $\phi\in\{\mathcal{H}\big|^{(1)},\delta c\}$, for this case and finds that the only nonvanishing commutation relations given by the Dirac brackets read
\begin{align}
\{  \delta q_i(\vec{k}),\delta \pi^j(-\vec{l})\}_D=\delta_i^{~j}\delta_{\vec{k},\vec{l}},~~\{ \delta \phi(\vec{k}),\delta p^{\phi}(-\vec{l})\}_D=\delta_{\vec{k},\vec{l}},
\end{align}
where $i,j=1,2,5,6$. Therefore, we reduce the phase space by replacing the weak equalities (\ref{vecH},\ref{vecG}) with the strong ones, $\delta q_3=\delta\pi_3=\delta q_4=\delta\pi_4=0$. Accordingly, the physical Hamiltonian does not depend on those variables and hence, they do not undergo any physical evolution.

In the kinematical phase space, the nonvanishing part of the second-order Hamiltonian that involves vector modes reads
\begin{align}\begin{split}
\mathcal{H}_{0}\big|^{(2V)}&=a[(\delta\pi^3)^2+(\delta\pi^4)^2]+\frac{a^{-1}p}{3}[(\delta\pi^3\delta q_3)+(\delta\pi^4\delta q_4)]\\ &+\left(\frac{5+3w}{72}a^{-3}p^2+\frac{a^{-3}k^2}{2}\right)[(\delta q_3)^2+(\delta q_4)^2],\end{split}
\end{align}
and clearly vanishes for the gauge-fixing conditions (\ref{vecG}). From the consistency equation (\ref{stabilityEq}),
\begin{equation}
\{\delta c_{\vec{v}},{\bf H}\} \approx 0,~~ \{\delta c_{\vec{w}},{\bf H}\}\approx 0,
\end{equation}
we obtain
\begin{align}
\delta N^{\vec{v}}\approx 0\approx \delta N^{\vec{w}},
\end{align}
and hence, all the vector modes of the four-dimensional metric perturbation must vanish.

\subsection{Dynamics of tensor modes}

Let us notice that the canonical pairs $(\delta q_5,\delta\pi_5)$ and $(\delta q_6,\delta\pi_6)$ are not constrained neither by the diffeomorphism constraints (\ref{diffeo}) nor by the scalar constraint (\ref{scalar}). The second-order Hamiltonian that involves tensor modes reads
\begin{align}\begin{split}
\mathcal{H}_{0}\big|^{(2T)}&=a[(\delta\pi^5)^2+(\delta\pi^6)^2]+\frac{a^{-1}p}{3}[(\delta\pi^5\delta q_5)+(\delta\pi^6\delta q_6)]\\
&+\left(\frac{5+3w}{72}a^{-3}p^2+\frac{a^{-3}k^2}{4}\right)[(\delta q_5)^2+(\delta q_6)^2].\end{split}
\end{align}
The tensor modes decouple from the remaining variables.

\subsection{Dynamics of scalar modes}

In what follows, we focus on the imposition of the remaining diffeomorphism constraint, namely $\mathcal{H}_{\vec{k}}\big|^{(1)}\approx 0$, and the scalar constraint, $\mathcal{H}_{0}\big|^{(1)}\approx 0$. We work with the variables that describe the scalar modes, $(\delta q_1,\pi^1,\delta q_2,\pi^2,\delta\phi,\delta p^{\phi})$. For the case of scalar modes, the Dirac procedure proves its merits most visibly. 

For each of the constraints $\mathcal{H}_{B}\big|^{(1)}\approx 0$, $B\in\{0,\vec{k}\}$, one chooses first-order {\it gauge fixing conditions}, denoted by $\delta c_{A}= 0$, $A\in\{0,\vec{k}\}$, such that the matrix
\begin{align}\label{LAZ}
\Lambda_{AB}=\left\{\delta c_{A},\mathcal{H}_{B}\big|^{(1)}\right\},
\end{align}
is invertible (this is equivalent to the condition (\ref{invB})). Then the Dirac brackets are introduced
\begin{align}\label{DB}
\{\cdot,\cdot\}_D=\{\cdot,\cdot\}-\{\cdot,\phi_{A}\}\{\phi_{A},\phi_{A'}\}^{-1}\{\phi_{A'},\cdot\},~~A,A'=0,\vec{k},~\phi=\mathcal{H}\big|^{(1)},\delta c.
\end{align}
Next, the weak equalities are replaced by the strong ones,
\begin{align}
\mathcal{H}_{A}\big|^{(1)}= 0~~\textrm{and}~~\delta c_{A}=0, ~~\textrm{where}~A\in\{0,\vec{k}\}.
\end{align}
Making use of the above relations, one can reduce the number of canonical variables in the formalism. In the process the initial Hamiltonian reduces to the physical one. If available, and this idea was developed in the context of canonical relativity by Kuchar \cite{doi:10.1063/1.1666050}, a convenient way to perform this reduction is to solve the first-class constraints $\mathcal{H}_{A}\big|^{(1)}=0$, $A\in\{0,\vec{k}\}$, with respect to two variables, say $\delta\pi_{n_1}$ and $\delta\pi_{n_2}$, while solving the gauge-fixing conditions $\delta c_{A}=0$, $A\in\{0,\vec{k}\}$, with respect to the canonically conjugate variables, $\delta q_{n_1}$ and $\delta q_{n_2}$. In this way it is quite straightforward to define the Dirac brackets by simply removing those variables from the definition of the Poisson brackets. This idea, however, works only for a limited number of gauge-fixing conditions.

In order to reconstruct the four-dimensional metric we need to study the consistency condition (\ref{stabilityEq}) which in this case reads
\begin{equation}\label{nfix}
\{\delta c_{A},{\bf H}\}=N\left\{\delta c_{A},\mathcal{H}_{0}\big|^{(0)}\right\}+N\int \left\{\delta c_{A},\mathcal{H}_{0}\big|^{(2S)}\right\}+\int\delta N^{B}\Lambda_{AB} \approx 0.
\end{equation}
Note that the consistency condition is first-order (we neglect higher orders). Once the gauge-fixing conditions $\delta c_{A}= 0$ are chosen, the above condition determines $\delta N^{B}$. Alternatively, in the context of the configuration space approach to cosmological perturbation theory one may and one commonly does impose conditions on $\delta N^{B}$. In this case, the consistency condition (\ref{nfix}) becomes a differential equation for the gauge-fixing conditions $\delta c_{A}$. Of course,  the solution to this differential equation is not unique and it depends on the initial condition. In other words, the choice of  $\delta N^{B}$ does not fix the {\it slicing} and the {\it threading} completely, which leads to the residual gauge freedom. A well-known example is the so-called synchronous gauge. In the next section we demonstrate with some examples both the first case and the mixed case that includes a single gauge-fixing condition $\delta c_{A}= 0$ and a single condition on $\delta N^{B}$.

The last issue we shall address is how to switch between different gauge-fixing conditions. By gauge-invariant variables in the standard cosmological perturbation theory one means first-order quantities whose Lie derivative along {\it any }first-order space-time vector field $\delta \vec{\xi}$ vanishes at first order. In the canonical framework, the gauge-invariant quantity $\delta D$ is the Dirac observable defined as (now we work again with the Poisson brackets in the kinematical phase space)
\begin{align}\label{deltaQ}
\int\delta\xi^{B}\left\{\mathcal{H}_{B}\big|^{(1)},\delta D\right\}\approx 0.
\end{align}
which holds up to first order. Provided that $\delta D=a_1\delta q_1+a_2\delta q_2+b_1\delta\pi_1+b_2\delta\pi_2+f_1\delta\phi+f_2\delta p^{\phi}$, then Eq. (\ref{deltaQ}) implies that the vector $(a_1,a_2,b_1,b_2,f_1,f_2)\in\mathbb{R}^6$ is orthogonal to
\begin{align}
\vec{v}_1&=\left(\frac{2}{3}a^2,2a^2,\frac{p}{3},-\frac{4p}{9},0,p^{\phi}\right)\\
\vec{v}_2&=\left(\frac{ap}{3},0,-\frac{a^{-1}p^2}{12}-2a^{-1}k^2-\frac{3}{2\alpha}\frac{a^{\frac{-3}{\alpha-1}}p^{T}}{a^2},\frac{2a^{-1}k^2}{3},-\frac{a^{\frac{-3}{\alpha-1}}p^{T}}{p^{\phi}},0\right).
\end{align}
We form the matrix $M=(\vec{v}_1,\vec{v}_2,\vec{0},\vec{0},\vec{0},\vec{0})$ made of $\vec{v}_1$, $\vec{v}_2$ and 4 vanishing vectors. The 4-dimendional null space of $M$ is the space of Dirac observables,
\begin{align}\label{D1}
\delta D_1&=\frac{\delta p^{\phi}}{\bar{p}^{\phi}}-\frac{\delta q_2}{2a^2},\\ \label{D2}
\delta D_2&=\bar{p}^{\phi}\delta\phi+\frac{3a^{\frac{-3}{\alpha-1}}p^{T}}{ap}\delta q_1-\frac{a^{\frac{-3}{\alpha-1}}p^{T}}{ap}\delta q_2,\\ \label{D3}
\delta D_3&=\delta\pi_1+\left(\frac{a^{-2}p}{4}+\frac{6a^{-2}k^2}{p}+\frac{9}{2\alpha}\frac{a^{-3}a^{\frac{-3}{\alpha-1}}p^{T}}{p}\right)\delta q_1-\left(\frac{a^{-2}p}{4}+\frac{2a^{-2}k^2}{p}+\frac{3}{2\alpha}\frac{a^{-3}a^{\frac{-3}{\alpha-1}}p^{T}}{p}\right)\delta q_2,\\ \label{D4}
\delta D_4&=\delta\pi_2-\frac{2a^{-2}k^2}{p}\delta q_1+\left(\frac{2a^{-2}p}{9}+\frac{2a^{-2}k^2}{3p}\right)\delta q_2.
\end{align}
Any combination of the above quantities multiplied by zero-order coefficients is a Dirac observable. Notice that their values are physically relevant only in the constraint surface and that the constraint functions are Dirac observables too. Indeed,
\begin{align}\begin{split}
\mathcal{H}_{0}\big|^{(1)}&=a^{\frac{-3}{\alpha-1}}p^{T}\delta D_1-\frac{ap}{3}\delta D_3+\frac{1}{2}\mathcal{H}_{0}\big|^{(0)}\delta q_2\approx a^{\frac{-3}{\alpha-1}}p^{T}\delta D_1-\frac{ap}{3}\delta D_3,\\
i\mathcal{H}_{\vec{k}}\big|^{(1)}&=\frac{2a^2}{3}\delta D_3+2a^2\delta D_4-\delta D_2+a^{-1}p^{-1}\mathcal{H}_{0}\big|^{(0)}(3\delta q_1-\delta q_2)\approx \frac{2a^2}{3}\delta D_3+2a^2\delta D_4-\delta D_2,\end{split}
\end{align}
where the above weak equalities mean ``in the zero-order constraint surface". Hence, there are only two independent Dirac observables. Let us denote them by $\Phi$ and $\Pi$. We define them on $\mathcal{H}_{0}\big|^{(0)}=0$ surface as follows:
\begin{align}\begin{split}\label{GIQ1}
\Phi&:=\delta D_2=\bar{p}^{\phi}\delta\phi+\frac{\alpha}{\alpha-1}\left(\frac{p}{6}\right)(3\delta q_1-\delta q_2),\\
\Pi&:=\delta D_1-\frac{3(\alpha-1)}{2}a^{-2}p^{-1}\delta D_2=\frac{\delta p^{\phi}}{\bar{p}^{\phi}}-\frac{3(\alpha-1)}{2}a^{-2}p^{-1}\bar{p}^{\phi}\delta\phi-\frac{3\alpha}{4}\frac{\delta q_1}{a^2}+\frac{\alpha-2}{4}\frac{\delta q_2}{a^2}.\end{split}
\end{align}
One may verify that $\{\Phi(\vec{k}) ,\Pi(-\vec{l})\}= \delta_{\vec{k},\vec{l}}$~  up to first order. Their physical interpretation is very ambiguous as it must rely on the kinematical phase space variables. Their physical meaning will be proposed upon choosing specific gauge-fixing conditions.

Suppose now that we are given two sets of gauge-fixing conditions (Fig. \ref{fig1}), 
\begin{align}
\delta c_1= 0= \delta c_2~~\textrm{and}~~\delta c'_1= 0= \delta c'_2,
\end{align}
and suppose that the physical dynamics confined to the surface given by the first set of conditions, $\delta c_1= 0 = \delta c_2$, is known to us. Now we wish to define the dynamics in the other surface which is given by the other set of conditions, $\delta c'_1= 0= \delta c'_2$. We express the dynamics in the first surface by means of the variables which are identical with the Dirac observables (\ref{GIQ1}) in that surface, namely $\Phi|_{\delta c_1=0=\delta c_2}$ and $\Pi|_{\delta c_1=0=\delta c_2}$. They form a complete set of variables in the reduced phase space. Next, using the basic canonical isomorphism we replace them with the same Dirac observables in the other gauge-fixing surface, namely
\begin{align}\label{GFCS}
\Phi|_{\delta c_1=0=\delta c_2}\mapsto \Phi|_{\delta c'_1=0=\delta c'_2},~~\Pi|_{\delta c_1=0=\delta c_2}\mapsto \Pi|_{\delta c'_1=0= \delta c'_2},
\end{align}
where
\begin{align}
\{\Phi|_{\delta c_1=0=\delta c_2},\Pi|_{\delta c_1=0=\delta c_2}\}_D=
\{\Phi|_{\delta c'_1=0=\delta c'_2}, \Pi|_{\delta c'_1=0= \delta c'_2}\}_{D'}.
\end{align}
In this way, the physical Hamiltonian and the respective dynamical equations for  $\Phi|_{\delta c'_1=0=\delta c'_2}$ and $\Pi|_{\delta c'_1=0= \delta c'_2}$ are obtained. Notice that this procedure makes sense only for the canonical variables and not for the Lagrange multipliers, $\delta N$ and $\delta N^{\vec{k}}:=i\vec{k}\delta n$. They have to be determined independently for each set of gauge-fixing conditions via Eq. (\ref{nfix}) in the kinematical phase space. Obviously, the domain of the definitions of $\delta N$ and $\delta n$ is the surface subject to the gauge-fixing conditions and thus, they can be also expressed in terms of the Dirac observables (\ref{GIQ1}).

\section{Examples of gauges}
We assume $K\alpha=1$ and introduce $p^T:=|\bar{p}^{\phi}|^{\frac{\alpha}{\alpha-1}}$. Recall that $\delta N^{\vec{k}}=i\vec{k}\delta n$.

\subsection{Spatially flat slicing}
In this gauge the constant time slices are assumed flat. Therefore, we impose the following gauge-fixing conditions,
\begin{align}\label{GFCSFS} 
\delta c_1:=\delta q_1= 0,~~\delta c_2:=\delta q_2= 0. 
\end{align}
We compute the matrix (\ref{LAZ}),
\begin{align}
\Lambda_{1\vec{k}}=i\frac{2a^2}{3},~~\Lambda_{10}=-\frac{ap}{3},~~\Lambda_{2\vec{k}}=i2a^2,~~\Lambda_{20}=0,
\end{align}
which is in fact invertible as $\textrm{det}|\Lambda_{AB}|=-i\frac{2}{3}a^3p\neq 0$. If we solve $\mathcal{H}_{B}\big|^{(1)}\approx 0$, $B\in\{0,\vec{k}\}$, with respect to $\delta\pi_1$ and $\delta\pi_2$, then it is straightforward to express the Dirac brackets (\ref{DB}) as
\begin{align}
\{ \delta \phi(\vec{k}),\delta p^{\phi}(-\vec{l})\}_D=\delta_{\vec{k},\vec{l}}~,
\end{align}
where we omit the tensor part of the brackets.

In the reduced phase space we introduce the reduced Hamiltonian by making use of the conditions $\delta c_1=\delta c_2=0=\mathcal{H}_{\vec{k}}\big|^{(1)}=\mathcal{H}_{0}\big|^{(1)}$ for reducing $\mathcal{H}_{0}\big|^{(2S)}$. We obtain
\begin{align}\label{preH}\begin{split}
&{\bf H}=N\mathcal{H}_{0}\big|^{(0)}+\\
&N\int\left( \frac{\alpha ap^2}{12(\alpha-1)^2}\left(\frac{\delta p^{\phi}}{\bar{p}^{\phi}}\right)^2-\frac{\alpha a^{-1}p}{4(\alpha-1)}(\delta\phi)(\delta p^{\phi}) +\left[\frac{3a^{-3}}{8}+\frac{3(\alpha-1)}{\alpha}\frac{a^{-3}k^2}{p^2}\right](\bar{p}^{\phi}\delta\phi)^2\right),\end{split}
\end{align}
(notice that the zero-order term is not reduced as we are not interested here in reducing the background variables). Observe that the reduced Hamiltonian is very simple to obtain as the gauge-fixing conditions kill in particular the complicated intrinsic curvature term. We make a canonical transformation (we use the constancy of $\bar{p}^{\phi}$),
\begin{align}\label{CTSFS}
\ud\delta\phi\ud\delta p^{\phi}=\ud\Phi\ud\Pi-\ud t\ud\left[\frac{3}{16}(\alpha-2)a^{-3}\Phi^2\right],
\end{align}
where
\begin{align}
\Pi=\frac{\delta p^{\phi}}{\bar{p}^{\phi}}-\frac{3(\alpha-1)}{2}a^{-2}p^{-1}\bar{p}^{\phi}\delta\phi,~~~\Phi=\bar{p}^{\phi}\delta\phi.
\end{align} 
We arrive at the following form of the Hamiltonian,
\begin{align}\label{SSFS}
{\bf H}=N\mathcal{H}_{0}\big|^{(0)}+
N\int\left( \frac{\alpha ap^2}{12(\alpha-1)^2}\Pi^2 +\frac{3(\alpha-1)}{\alpha}\frac{a^{-3}k^2}{p^2}\Phi^2\right).
\end{align}
Notice that the variables $\Phi$ and $\Pi$ correspond to the Dirac observables (\ref{GIQ1}) in the considered gauge and therefore we use the same symbols\footnote{Actually, for more precise notation one should use the symbols $\Phi|_{\delta c_1=0=\delta c_2}$ and $\Pi|_{\delta c_1=0=\delta c_2}$, respectively.}.  

The Hamilton equations for the gauge-invariant perturbation variables in conformal time ($N=a$) read:
\begin{align}\label{HEMsfs}
\acute{\Phi}= \frac{\alpha a^2p^2}{6(\alpha-1)^2}\Pi,~~\acute{\Pi}=-\frac{6(\alpha-1)}{\alpha}\frac{a^{-2}k^2}{p^2}\Phi.
\end{align}
Later on, we will confirm this result by comparing it with the results previously known in the literature. Now, we compute $\delta N$ and $\delta N^{i}=i\delta nk^i$. Eq. (\ref{nfix}) yields
\begin{align}
N\left(-\frac{1}{3}a\delta\pi_1-\frac{1}{6}a^{-1}p\delta q_1\right)-\delta N\frac{ap}{3}-\delta n k^2\frac{2a^2}{3}=0,\\
N\left(3a\delta\pi_2+\frac{1}{3}a^{-1}p\delta q_2\right)-\delta n k^22a^2=0,
\end{align}
which can be easily solved and the solution can be expressed in terms of $\bar{p}^{\phi}\delta\phi$ and $\frac{\delta p^{\phi}}{\bar{p}^{\phi}}$,
\begin{align}\label{metricSFS}
\frac{\delta N}{N}=-\frac{3}{2}a^{-2}p^{-1}\bar{p}^{\phi}\delta\phi,~~\frac{\delta n}{N}=\frac{3}{4}a^{-3}k^{-2}\bar{p}^{\phi}\delta\phi-\frac{3}{2}a^{-3}p^{-1}k^{-2}a^{\frac{-3}{\alpha-1}}p^{T}\frac{\delta p^{\phi}}{\bar{p}^{\phi}}.
\end{align}
This completes the canonical formulation of the dynamics of the scalar perturbation in the spatially flat slicing gauge. We give the physical meaning to the Dirac observables $\Phi$ and $\Pi$ for the present and other choices of gauge-fixing conditions in Appendix \ref{App1}.

\subsection{Uniform density} 
Above we have computed the dynamics of the scalar perturbation in the spatially flat slicing gauge in terms of the perturbation variables, which on the gauge fixing surface $\delta c_1= 0$ and $\delta c_2= 0$ of Eq. (\ref{GFCSFS}), are identical to the Dirac observables, $\Phi$ and $\Pi$, defined in Eqs (\ref{GIQ1}). This result allows us to switch swiftly from the spatially flat slicing gauge to any other gauge. We call the present gauge the {\it uniform density gauge} as it assumes the perturbation of the energy density of matter to vanish at the constant time slices. 

The first-order density reads
\begin{align}
\delta\rho=\frac{\mathcal{H}_{f,0}}{\sqrt{q}}\bigg|^{(1)}=a^{-3}a^{\frac{-3}{\alpha-1}}p^T\left(\frac{\delta p^{\phi}}{\bar{p}^{\phi}}-\frac{3 \delta q_1}{2a^{2}}\right).
\end{align}
The condition for the vanishing of the density perturbation $\delta\rho$ constitutes a partial gauge-fixing condition. We complement it with the condition of vanishing of the perturbation $\delta p^{\phi}$. Specifically, we impose the following gauge-fixing conditions,
\begin{align}
\delta c_1=a^{\frac{-3}{\alpha-1}}p^T\left(\frac{\delta p^{\phi}}{\bar{p}^{\phi}}-\frac{3 \delta q_1}{2a^{2}}\right),~~\delta c_2=a^{\frac{-3}{\alpha-1}}p^T\frac{\delta p^{\phi}}{\bar{p}^{\phi}}.
\end{align}
We compute the matrix (\ref{LAZ}),
\begin{align}
\Lambda_{1\vec{k}}=0,~~\Lambda_{10}=a^{\frac{-3}{\alpha-1}}p^T\frac{a^{-1}p}{2},~~\Lambda_{2\vec{k}}=-ia^{\frac{-3}{\alpha-1}}p^T,~~\Lambda_{20}=0,
\end{align}
which is in fact invertible as $\textrm{det}|\Lambda_{AB}|=i\frac{a^{-1}p}{2}(a^{\frac{-3}{\alpha-1}}p^T)^2\neq 0$. Given these conditions, the Dirac observables (\ref{GIQ1}) read
\begin{align}
\Phi&=\bar{p}^{\phi}\delta\phi-\frac{\alpha}{\alpha-1}\left(\frac{p}{6}\right)\delta q_2,\\
\Pi&=-\frac{3(\alpha-1)}{2}a^{-2}p^{-1}\bar{p}^{\phi}\delta\phi+\frac{\alpha-2}{4}\frac{\delta q_2}{a^2}.
\end{align}

The dynamics of the above variables is given again by the Hamiltonian (\ref{SSFS}), i.e.,
\begin{align}
{\bf H}=N\mathcal{H}_{0}\big|^{(0)}+
N\int\left( \frac{\alpha ap^2}{12(\alpha-1)^2}\Pi^2 +\frac{3(\alpha-1)}{\alpha}\frac{a^{-3}k^2}{p^2}\Phi^2\right).
\end{align}

Now, we compute $\delta N$ and $\delta N^{i}=i\delta nk^i$. Eq. (\ref{nfix}) yields
\begin{align}
Na^{\frac{-3}{\alpha-1}}p^T\left(\frac{1}{2a}\delta\pi_1-\frac{\alpha+2}{4(\alpha-1)}\frac{p\delta q_1}{a^2}-\frac{k^2}{a^2}\bar{p}^{\phi}\delta\phi+\frac{a^{-1}p}{2(\alpha-1)}\frac{\delta p^{\phi}}{\bar{p}^{\phi}}\right)+\delta N\frac{a^{-1}p}{2}a^{\frac{-3}{\alpha-1}}p^T=0,\\
N\left(-\frac{k^2}{a^2}\bar{p}^{\phi}\delta\phi+\frac{a^{-1}p}{2(\alpha-1)}a^{\frac{-3}{\alpha-1}}p^T\frac{\delta p^{\phi}}{\bar{p}^{\phi}}\right)-\delta n k^2a^{\frac{-3}{\alpha-1}}p^T=0,
\end{align}
which has the solution in the gauge-fixing surface,
\begin{align}\label{ugls}
\frac{\delta n}{N}&=\frac{a^{-2}}{a^{\frac{-3}{\alpha-1}}p^{T}}\left(-\bar{p}^{\phi}\delta\phi+\frac{ap}{2k^2}\frac{a^{\frac{-3}{\alpha-1}}p^{T}}{\alpha-1}\frac{\delta p^{\phi}}{\bar{p}^{\phi}}\right),\\ \nonumber
\frac{\delta N}{N}&=\frac{a^{-1}p^{-1}}{a^{\frac{-3}{\alpha-1}}p^{T}}\left(2k^2a^{\frac{-3}{\alpha-1}}p^{T}\bar{p}^{\phi}\delta\phi-a^{\frac{-3}{\alpha-1}}p^{T}a\delta\pi_1\right).
\end{align}

\subsection{Comoving orthogonal}
The comoving orthogonal gauge assumes the flow of the fluid orthogonal to constant time slices, the condition that is expressed by Eq. (\ref{orthogonal}). In other words, the perturbation of the fluid variable $\delta\phi$ vanishes at the constant time slices in this gauge, i.e. $\delta c_1:=\delta\phi= 0$, which is the first gauge-fixing condition. We are going to set $\delta c_2:=\delta p^{\phi}= 0$ as the other gauge-fixing condition. We will see that they are indeed consistent with each other.

Therefore, let us set
\begin{align}\label{COgfc}
\delta c_1:=\delta\phi= 0,~~\delta c_2:=\delta p^{\phi}= 0.
\end{align} 
We compute the matrix (\ref{LAZ}),
\begin{align}
\Lambda_{1\vec{k}}=0,~~\Lambda_{10}=\frac{a^{\frac{-3}{\alpha-1}}p^T}{\bar{p}^{\phi}},~~\Lambda_{2\vec{k}}=-i\bar{p}^{\phi},~~\Lambda_{20}=0,
\end{align}
which is in fact invertible as $\textrm{det}|\Lambda_{AB}|=ia^{\frac{-3}{\alpha-1}}p^T\neq 0$. Given these conditions, the Dirac observables (\ref{GIQ1}) read
\begin{align}
\Phi&=\frac{\alpha}{\alpha-1}\left(\frac{p}{6}\right)(3\delta q_1-\delta q_2),\\
\Pi&=-\frac{3\alpha}{4}\frac{\delta q_1}{a^2}+\frac{\alpha-2}{4}\frac{\delta q_2}{a^2}.
\end{align}

The dynamics of these variables is given again by the Hamiltonian (\ref{SSFS}), i.e.,
\begin{align}
{\bf H}=N\mathcal{H}_{0}\big|^{(0)}+
N\int\left( \frac{\alpha ap^2}{12(\alpha-1)^2}\Pi^2 +\frac{3(\alpha-1)}{\alpha}\frac{a^{-3}k^2}{p^2}\Phi^2\right).
\end{align}

We determine $\delta N$ and $\delta N^{i}=i\delta nk^i$. Eq. (\ref{nfix}) yields
\begin{align}\label{COls1}
N\left(\frac{a^{\frac{-3}{\alpha-1}}p^T}{\alpha-1}\frac{\delta p^{\phi}}{(\bar{p}^{\phi})^2}-\frac{3a^{\frac{-3}{\alpha-1}}p^T}{2(\alpha-1)}\frac{\delta q_1}{a^2\bar{p}^{\phi}}\right)+\delta N\frac{a^{\frac{-3}{\alpha-1}}p^T}{\bar{p}^{\phi}}=0,\\ \label{COls2}
N\left(-k^2a\mu^{\alpha-2}\delta\phi\right)+\delta n k^2\bar{p}^{\phi}=0,
\end{align}
and the solution in the gauge-fixing surface reads,
\begin{align}
\delta n =0,~~\frac{\delta N}{N}=
\frac{3\delta q_1}{2a^2(\alpha-1)}.
\end{align}
In other words, the threading is also orthogonal to constant time slices as $\delta n\approx 0$.  

\subsection{Longitudinal} 
The longitudinal gauge assumes $\delta q_2=0$ and $\delta n =0$ \cite{Malik:2008im}. This is the mixed case that we announced in Sec. IV. From the Hamiltonian point of view there is only one gauge-fixing condition, namely $\delta c_1:=\delta q_2= 0$, since $\delta n$ does not belong to the phase space. Hence, we need to propose another gauge-fixing condition, $\delta c_2= 0$. Let us first study the consistency condition (\ref{nfix}) assuming $\delta n =0$. We verify that $\delta c_1=0$ is consistent with the dynamics:
\begin{align}
\{\delta c_1,{\bf H}\} = N\left\{\delta q_{2},\mathcal{H}_{0}\big|^{(2S)}\right\}=N\left(3a\delta\pi_2+\frac{1}{3}a^{-1}p\delta q_2\right)= 3Na\delta\pi_2,
\end{align}
if we assume $\delta c_2:=\delta\pi_2= 0$ for consistency. We see that there is no ambiguity in defining the gauge-fixing surface contrary to our expectations expressed in Sec. IV. This seems unusual and follows from the fact that the first gauge-fixing condition, $\delta q_{2}=0$, commutes with the scalar constraint and as a result, the above consistency condition does not depend on $\delta N$ and unambiguously yields the second gauge-fixing condition. To confirm that this is indeed a valid choice of gauge, let us compute the matrix (\ref{LAZ}),
\begin{align}
\Lambda_{1\vec{k}}=i2a^2,~~\Lambda_{10}=0,~~\Lambda_{2\vec{k}}=-i\frac{4}{9}p,~~\Lambda_{20}=-\frac{2}{3}a^{-1}k^2,
\end{align}
which is in fact invertible as $\textrm{det}|\Lambda_{AB}|=-i\frac{4}{3}k^2a\neq 0$. Given these conditions, the Dirac observables (\ref{GIQ1}) read
\begin{align}
\Phi&=\bar{p}^{\phi}\delta\phi+\frac{\alpha}{\alpha-1}\left(\frac{p}{2}\right)\delta q_1,\\
\Pi&=\frac{\delta p^{\phi}}{\bar{p}^{\phi}}-\frac{3(\alpha-1)}{2}a^{-2}p^{-1}\bar{p}^{\phi}\delta\phi-\frac{3\alpha}{4}\frac{\delta q_1}{a^2}.
\end{align}

The dynamics of these variables is given again by the Hamiltonian (\ref{SSFS}), i.e.,
\begin{align}
{\bf H}=N\mathcal{H}_{0}\big|^{(0)}+
N\int\left( \frac{\alpha ap^2}{12(\alpha-1)^2}\Pi^2 +\frac{3(\alpha-1)}{\alpha}\frac{a^{-3}k^2}{p^2}\Phi^2\right).
\end{align}

Then, from the consistency condition (\ref{nfix}) for $\delta c_2=0$,
\begin{align}
\{\delta c_2,{\bf H}\} = N\left\{\delta\pi_{2},\mathcal{H}_{0}\big|^{(2S)}\right\}+\delta N\left\{\delta\pi_{2},\mathcal{H}_{0}\big|^{(1)}\right\}\approx -N\frac{1}{3}a^{-3}k^2\delta q_1-\delta N\frac{2}{3}a^{-1}k^2=0,
\end{align}
one finds
\begin{align}\label{gauge1}
\frac{\delta N}{N}=-\frac{1}{2}a^{-2}\delta q_1.
\end{align} 

Now, one can show that in the longitudinal gauge the Newtonian potential $\frac{\delta N}{N}$ in terms of gauge-invariant variables $\Phi$, $\Pi$ reads:
\begin{align}\label{gauge2}
\frac{\delta N}{N}=-\frac{(\alpha-2)p}{16a^2k^2}\Phi-\frac{\alpha p^2}{24(\alpha-1)k^2}\Pi,
\end{align}
Plugging the above into the equations of motion (\ref{HEMsfs}) in conformal time $N=a$ yields:
\begin{align}
\left(\frac{\delta N}{a}\right)''-\frac{\alpha}{2(\alpha-1)}p\left(\frac{\delta N}{a}\right)'+\frac{k^2}{\alpha-1}\left(\frac{\delta N}{a}\right)=0,
\end{align}
where $':=\frac{\partial}{\partial\eta}$. This result agrees with Eq. (12.9) of \cite{Mukhanov:1990me}. Notice that in order to arrive at this result we have taken a few steps: (1) expanded the ADM formalism around the flat FLRW model, (2) imposed the spatially flat slicing gauge-fixing conditions, (3) computed the respective Dirac brackets, (4) derived the Dirac observables and identified them with the variables used in the spatially flat gauge and (5) used the canonical isomorphism to deduce the dynamics of the perturbation in the longitudinal gauge. A mistake in any of those steps would almost certainly lead to a wrong result. Hence, the right result derived above confirms the validity of the {\it entire} formalism.

\section{Multi-fluid case}
In this section we give a generalisation of the above result to the multi-fluid case. { The multi-fluid case was previously studied in \cite{Peter:2015zaa}, where the authors derive the reduced phase space by solving the constraints with respect to some momenta and re-arranging the obtained terms in such a way that the unconstrained Hamiltonian can be deduced. The concepts of Dirac observables, Dirac brackets and the canonical isomorphism between gauge-fixing surfaces are not used in their analysis. 

While the gravitational part of the constraints remains unchanged,} the fluid part of the constraints is now simply extended to include many fluids labelled by $i$, each satisfying the equation of state $p_i=w_i\rho_i$,
\begin{align}
\mathcal{H}_{f,0}\big|^{(0)}=\sum_i(\alpha_i-1)a^3K_i\left(\frac{|\bar{p}^{\phi_i}|}{K_i\alpha_i a^3}\right)^{\frac{\alpha_i}{\alpha_i-1}},
\end{align}
\begin{align}
\mathcal{H}_{f,0}\big|^{(1)}=\sum_iK_i\alpha_i a^3\left(\frac{|\bar{p}^{\phi_i}|}{K_i\alpha_i a^3}\right)^{\frac{\alpha_i}{\alpha_i-1}}\left(\frac{\delta p^{\phi_i}}{\bar{p}^{\phi_i}}-\frac{3}{2\alpha_i}\frac{\delta q_1}{a^2}\right),
\end{align}
\begin{align}
\mathcal{H}_{f,a}\big|^{(1)}=\sum_i\bar{p}^{\phi_i}\delta\phi_{i,a},
\end{align}
\begin{align}\begin{split}
&\mathcal{H}_{f,0}\big|^{(2)}=\sum_iK_i\sqrt{q}\mu_i^{\alpha_i}\alpha_i\\
&\times\left(\frac{1}{2(\alpha_i-1)}\left(\frac{\delta p^{\phi_i}}{\bar{p}^{\phi_i}}\right)^2-\frac{3}{2(\alpha_i-1)}\left(\frac{\delta q_1}{a^2}\right)\left(\frac{\delta p^{\phi_i}}{\bar{p}^{\phi_i}}\right)+\frac{9}{8\alpha_i(\alpha_i-1)}\left(\frac{\delta q_1}{a^2}\right)^2+\frac{1}{4\alpha_i}\frac{3(\delta q_1)^2+\frac{2}{3}(\delta q_2)^2}{a^4}\right)\\ 
&~~~~~~~~~+k^2K_i\sqrt{q}\mu_i^{\alpha_i-2}\frac{\alpha_i}{2}a^{-2}\left(\delta\phi_i\right)^2,\end{split}
\end{align}
where $\alpha_i=\frac{w_i+1}{w_i}$. We will assume $K_i\alpha_i=1$ for all fluids. First, let us observe that in the case of many perfect fluids, the tensor and the vector parts of the second-order Hamiltonian and first-order constraints consist purely of the gravitational field perturbations. Therefore, the analysis of the tensor and vector modes is identical to the single fluid case, presented in Sec. \ref{IV}. In what follows we focus on the scalar perturbation.

It is very convenient to work in the spatially flat slicing gauge ($\delta q_1=0=\delta q_2$) and, in principle, repeat all the steps made for the single-fluid case. If we solve $\mathcal{H}_{B}\big|^{(1)}\approx 0$, $B\in\{0,\vec{k}\}$, with respect to $\delta\pi_1$ and $\delta\pi_2$, then it is straightforward to find the Dirac brackets (\ref{DB}) between the matter perturbation variables,
\begin{align}
\{ \delta \phi_i(\vec{k}),\delta p^{\phi_j}(-\vec{l})\}_D=\delta_i^{~j}\delta_{\vec{k},\vec{l}}~,
\end{align}
where we omit the tensor part of the brackets. Let us introduce 
\begin{align}\label{SFSginv}
\Phi_i:=\bar{p}^{\phi_i}\delta\phi_i,~~\Pi_i:=\frac{\delta p^{\phi_i}}{\bar{p}^{\phi_i}}.
\end{align} 
Notice that
\begin{align}
\{ \Phi_i(\vec{k}),\Pi_j(-\vec{l})\}_D=\delta_{ij}\delta_{\vec{k},\vec{l}}~,
\end{align}
and no extra term in the Hamiltonian is generated by this coordinate transformation as $\bar{p}^{\phi_i}$'s are constants of motion. The reduced Hamiltonian reads
\begin{align}\label{HamMF}\begin{split}
{\bf H}=N\mathcal{H}_{0}\big|^{(0)}+\sum_i \frac{a^{\frac{-3}{\alpha_i-1}}p^{T_i}}{2(\alpha_i-1)}\Pi_i^2\\
-\frac{3a^{-2}p^{-1}}{2}\left(\sum_k\Phi_k\right)\left(\sum_ia^{\frac{-3}{\alpha_i-1}}p^{T_i}\Pi_i\right)+\frac{3a^{-3}}{8}\left(\sum_k\Phi_k\right)^2+\sum_i\frac{k^2a^{-2}}{2a^{\frac{-3}{\alpha_i-1}}p^{T_i}}\Phi_i^2,\end{split}
\end{align}
where $p^{T_i}:=|\bar{p}^{\phi_i}|^{\frac{\alpha_i}{\alpha_i-1}}$. It can be readily verified that the above formula coincides with Eq. (\ref{preH}) for a single fluid. Notice that contrary to the single-fluid case, in the multi-fluid case there is no canonical transformation that can diagonalize the reduced Hamiltonian. In order to reconstruct the full space-time metric in this gauge, one solves the consistency equation (\ref{stabilityEq}):
\begin{align}\begin{split}
\{\delta q_{1},{\bf H}\}=N(-\frac{a}{3}\delta\pi_1-\frac{a^{-1}p}{6}\delta q_1)-\delta N(\frac{ap}{3})-\delta n (\frac{2}{3}a^2k^2)\approx 0,\\
\{\delta q_{2},{\bf H}\}=N(3a\delta\pi_2+\frac{a^{-1}p}{3}\delta q_2)-\delta n (2a^2k^2)\approx 0,\end{split}
\end{align}
which in the gauged-fixed surface $\delta q_1=\delta q_2=\mathcal{H}_{\vec{k}}\big|^{(1)}=\mathcal{H}_{0}\big|^{(1)}=0$ has the solution,
\begin{align}
\frac{\delta N}{N}=-\frac{3}{2}a^{-2}p^{-1}\sum_i\bar{p}^{\phi_i}\delta\phi_{i},~~\frac{\delta n}{N}=\frac{3}{4}a^{-3}k^{-2}\sum_i\bar{p}^{\phi_i}\delta\phi_{i}-\frac{3}{2}a^{-3}p^{-1}k^{-2}\sum_ia^{\frac{-3}{\alpha_i-1}}p^{T_i}\frac{\delta p^{\phi_i}}{\bar{p}^{\phi_i}},
\end{align}
which agrees with Eq. (\ref{metricSFS}) in the single fluid case.

\subsection{Switching between gauges}
In order to switch between gauges it is useful to define the Dirac observables analogously to the single-fluid case,
\begin{align}
\delta D=a_1\delta q_1+a_2\delta q_2+b_1\delta\pi_1+b_2\delta\pi_2+\sum_if_{1i}\delta\phi_i+\sum_if_{2i}\delta p^{\phi_i}.
\end{align}
The vectors $(a_1,a_2,b_1,b_2,f_{i1},f_{i2})\in\mathbb{R}^{4+2n}$, where $n$ is the number of fluids, are orthogonal to
\begin{align}
\vec{v}_1&=\left(\frac{2}{3}a^2,2a^2,\frac{p}{3},-\frac{4p}{9},0,p^{\phi_1},0,p^{\phi_2},\dots\right)\\
\vec{v}_2&=\left(\frac{ap}{3},0,-\frac{a^{-1}p^2}{12}+2a^{-1}k^2-\sum_i\frac{3}{2\alpha_i}\frac{a^{\frac{-3}{\alpha_i-1}}p^{T_i}}{a^2},-\frac{2a^{-1}k^2}{3},-\frac{a^{\frac{-3}{\alpha_1-1}}p^{T_1}}{p^{\phi_1}},0,-\frac{a^{\frac{-3}{\alpha_2-1}}p^{T_2}}{p^{\phi_2}},0,\dots\right).
\end{align}
We form the matrix $M=(\vec{v}_1,\vec{v}_2,\vec{0},\vec{0},\vec{0}_1,\vec{0}_1,\dots,\vec{0}_n,\vec{0}_n)$ made of $\vec{v}_1$, $\vec{v}_2$ and $2(n+1)$ null vectors. The $2(n+1)$-dim null space of $M$ is the space of Dirac observables spanned by
\begin{align}
\delta D_{1i}&=\frac{\delta p^{\phi_i}}{\bar{p}^{\phi_i}}-\frac{\delta q_2}{2a^2},\\
\delta D_{2i}&=\bar{p}^{\phi_i}\delta\phi_i+\frac{3a^{\frac{-3}{\alpha_i-1}}p^{T_i}}{ap}\delta q_1-\frac{a^{\frac{-3}{\alpha_i-1}}p^{T_i}}{ap}\delta q_2,\\
\delta D_3&=\delta\pi_1+\left(\frac{a^{-2}p}{4}+\frac{6a^{-2}k^2}{p}+\sum_i\frac{9}{2\alpha_i}\frac{a^{\frac{-3}{\alpha_i-1}}p^{T_i}}{a^{3}p}\right)\delta q_1-\left(\frac{a^{-2}p}{4}+\frac{2a^{-2}k^2}{p}+\sum_i\frac{3}{2\alpha_i}\frac{a^{\frac{-3}{\alpha_i-1}}p^{T}}{a^3p}\right)\delta q_2,\\
\delta D_4&=\delta\pi_2-\frac{2a^{-2}k^2}{p}\delta q_1+\left(\frac{2a^{-2}p}{9}+\frac{2a^{-2}k^2}{3p}\right)\delta q_2.
\end{align}
Any combination of the above quantities formed with zero-order coefficients is a Dirac observable. Notice that their values are physically relevant  only in the constraint surface and that the constraint functions are Dirac observables too. Indeed,
\begin{align}
\mathcal{H}_{0}\big|^{(1)}&=\sum_ia^{\frac{-3}{\alpha_i-1}}p^{T_i}\delta D_{1i}-\frac{ap}{3}\delta D_3+\frac{1}{2}\mathcal{H}_{0}\big|^{(0)}\delta q_2,\\
i\mathcal{H}_{\vec{k}}\big|^{(1)}&=\frac{2a^2}{3}\delta D_3+2a^2\delta D_4-\sum_i\delta D_{2i}+a^{-1}p^{-1}\mathcal{H}_{0}\big|^{(0)}(3\delta q_1-\delta q_2).
\end{align}
Therefore, there are $2n$  independent Dirac observables. As in the single fluid case, we define the following set of independent Dirac observables:
\begin{align}\label{multiDirac}\begin{split}
\Phi_i&:=\delta D_{2i}=\bar{p}^{\phi_i}\delta\phi_i+\frac{a^{\frac{-3}{\alpha_i-1}}p^{T_i}}{ap}\left(3\delta q_1-\delta q_2\right),\\
\Pi_i&:=\delta D_{1i}=\frac{\delta p^{\phi_i}}{\bar{p}^{\phi_i}}-\frac{\delta q_2}{2a^2}.\end{split}
\end{align}
One may verify up to first order that $\{\Phi_i(\vec{k}) ,\Pi_j(-\vec{l})\}= \delta_{ij}\delta_{\vec{k},\vec{l}}$ with all other commutation relations vanishing. The reduced Hamiltonian (\ref{HamMF}) is now interpreted as physical, i.e. expressed in gauge-invariant variables, $\Phi_i$ and $\Pi_i$, defined above, which in the spatially flat slicing gauge coincide with $\Phi_i$ and $\Pi_i$ of Eq. (\ref{SFSginv}).

Now, as in the case of a single fluid, we may consider {\it any} gauge-fixing conditions. For example, the {\it uniform density gauge} assumes that there is no energy density perturbation at the constant time slices, i.e.
\begin{align}
\delta c_1:=a^3\delta\rho=a^3\frac{\mathcal{H}_{f,0}}{\sqrt{q}}\bigg|^{(1)}=\sum_ia^{\frac{-3}{\alpha_i-1}}p^{T_i}\left(\frac{\delta p^{\phi_i}}{\bar{p}^{\phi_i}}-\frac{3 \delta q_1}{2a^{2}}\right)=0.
\end{align}
We may supplement the above condition with
\begin{align}
\delta c_2:=\sum_ia^{\frac{-3}{\alpha_i-1}}p^{T_i}\frac{\delta p^{\phi_i}}{\bar{p}^{\phi_i}}=0.
\end{align}
In order to reconstruct the full space-time metric in a given gauge, one solves the consistency equation (\ref{stabilityEq}):
\begin{equation}
\{\delta c_{i},{\bf H}\} \approx 0,~~i=1,2,
\end{equation}
from which we obtain
\begin{align}
\frac{\delta n}{N}&=\frac{a^{-2}}{\sum_ia^{\frac{-3}{\alpha_i-1}}p^{T_i}}\left(-\sum_i\bar{p}^{\phi_i}\delta\phi_i+\frac{ap}{2k^2}\sum_i\frac{a^{\frac{-3}{\alpha_i-1}}p^{T_i}}{\alpha_i-1}\frac{\delta p^{\phi_i}}{\bar{p}^{\phi_i}}\right),\\
\frac{\delta N}{N}&=\frac{a^{-1}p^{-1}}{\sum_ia^{\frac{-3}{\alpha_i-1}}p^{T_i}}\left(2k^2\sum_ia^{\frac{-3}{\alpha_i-1}}p^{T_i}\bar{p}^{\phi_i}\delta\phi_i-(\sum_ia^{\frac{-3}{\alpha_i-1}}p^{T_i})a\delta\pi_1-ap\sum_i\frac{a^{\frac{-3}{\alpha_i-1}}p^{T_i}}{\alpha_i-1}\frac{\delta p^{\phi_i}}{\bar{p}^{\phi_i}}\right),
\end{align}
which agrees with Eq. (\ref{ugls}) in the single fluid case.

\section{Conclusion}
In the present paper we have derived the reduced phase space formulation of cosmological perturbation theory by applying the standard Dirac procedure to the ADM formalism expanded around the FLRW universe. {  The concepts of Dirac observables,  Dirac brackets and the canonical isomorphism between gauge-fixing surfaces are the basic means of our derivation.} Our formalism accommodates all choices of gauge-fixing conditions and they are related by the mentioned canonical isomorphisms. The gauge-fixing surfaces can be related to the choices of gauges made in the standard configuration space approach. Our formalism includes a prescription for the reconstruction of the four-dimensional metric and is easily extended to the multi-fluid case.

{ 
The current challenge is to extend the cosmological perturbation theory to higher orders. Already at second order the definitions of gauge-invariant variables and gauge-fixing conditions become cumbersome in the standard approach (see e.g. \cite{Uggla:2019}). The present approach is in principle applicable to {\it any} order and provides a valid alternative proposal for tackling this difficult task. In our opinion, our approach is computationally effective and conceptually clear, and it automatically endows the perturbation theory with the right symplectic structure needed for quantization.  It is worth noting that there are more alternative approaches that are being developed (see e.g. \cite{Giesel:2018opa}) with an eye towards higher-order cosmological perturbation theory and its quantization. 

The canonical language of our approach is often used in canonical relativity. One of its advantages is that it establishes a basic link between quantizations of the kinematical phase space (``Dirac quantizations” made before solving the constraints such as the Wheeler-DeWitt equation), and the reduced phase space quantizations (made after solving the constraints). Indeed, our approach provides a geometrical relation between the reduced phase space and the kinematical phase space. As some canonical approaches to quantum gravity are based on the Dirac quantization while others on the reduced phase space quantization, the present work provides a tool for their comparisons. 

The obtained perturbed Hamiltonian has in our opinion several advantages over the standard Hamiltonian given in Eq. (10.62) of \cite{Mukhanov:1990me} based on the Mukhanov-Sasaki variable if the background spacetime is also to be quantized. Note that all the background quantities in the perturbed Hamiltonian (\ref{SSFS}) are now expressed purely in terms of canonical variables and hence they can be quantized directly at this level. One the other hand, the background quantities in the Hamiltonian (10.62) of \cite{Mukhanov:1990me} are made exclusively of the configuration variable, the scale factor. When one quantizes the background, the direct approach results in diffferent quantum dynamics of perturbations than the mere substitution of the classical scale factor and its derivatives with an expectation value of the quantized scale factor and its derivatives. For example, in case of the radiation-filled universe the direct approach gives the vanishing of the amplification potential $U$ of Eq. (\ref{potential}), whereas the configuration space based approach gives a nontrivial $U$.  As another advantage, in the present approach the definitions of Dirac observables and of the Hamiltonian do not rely on any particular choice of the background time such as conformal time. The freedom in choice of background time is useful. For instance, whether quantisation of the background spacetime resolves the initial singularity may depend on the internal time employed in defining the reduced phase space. Hence, the quantum dynamics of perturbation also depends on this choice and our approach provides a means to study this dependence.

Finally, it is worth noting that the gauge-fixing procedure used in our approach may be used as a convenient tool for exploring new gauges: first, it gives a clear distinction between gauges with residual gauge freedom (involving conditions for lapse and shift which are not phase space variables) and complete gauges (involving only phase space variables); second, it gives a clear test of the validity of any gauge (i.e. the non-vanishing of the determinant (\ref{invB})).}

\acknowledgments This work is a part of the research project 2018/30/E/ST2/00370 financed by National Science Centre, Poland. I thank Jean-Pierre Gazeau (APC, Universit\'e Paris Diderot) for his helpful comments on an earlier version of the manuscript.

\appendix

\section{Expansion of the ADM Hamiltonian}\label{expansion}
The total Hamiltonian expanded up to second order reads
\begin{equation}
{\bf H}=N\mathcal{H}_{0}\big|^{(0)}+\int (N\mathcal{H}_{0}\big|^{(2)}+\delta N\mathcal{H}_{0}\big|^{(1)}+\delta N^a\mathcal{H}_{a}\big|^{(1)})~\ud^3x,
\end{equation}
where
$\mathcal{H}_{0}\big|^{(0)}=\mathcal{H}_{g,0}\big|^{(0)}+\mathcal{H}_{f,0}\big|^{(0)}$,
\begin{align}
\mathcal{H}_{g,0}\big|^{(0)}=-\frac{1}{6}ap^2,~~\mathcal{H}_{f,0}\big|^{(0)}=(\alpha-1)a^3K\left(\frac{|\bar{p}^{\phi}|}{K\alpha a^3}\right)^{\frac{\alpha}{\alpha-1}},
\end{align}
and  $\mathcal{H}_{0}\big|^{(1)}=\mathcal{H}_{g,0}\big|^{(1)}+\mathcal{H}_{f,0}\big|^{(1)}$,
\begin{align}\begin{split}\label{con10}
\mathcal{H}_{g,0}\big|^{(1)}&=-\frac{ap}{3}\delta\pi-\frac{a^{-1}p^2}{36}\delta q-a^{-1}(\delta q_{ab,ab}-\delta q_{aa,bb}),\\ 
\mathcal{H}_{f,0}\big|^{(1)}&=K\alpha a^3\left(\frac{|\bar{p}^{\phi}|}{K\alpha a^3}\right)^{\frac{\alpha}{\alpha-1}}\left(\frac{\delta p^{\phi}}{\bar{p}^{\phi}}-\frac{1}{2\alpha}\frac{\delta q}{a^2}\right),\end{split}
\end{align}
and $\mathcal{H}_{a}\big|^{(1)}=\mathcal{H}_{g,a}\big|^{(1)}+\mathcal{H}_{f,a}\big|^{(1)}$,
\begin{align}\label{con1a}
\mathcal{H}_{g,a}\big|^{(1)}=-2(a^2\delta\pi^{ab}_{,b}+\frac{1}{3}p\delta q_{ab,b}),~~ 
\mathcal{H}_{f,a}\big|^{(1)}=\bar{p}^{\phi}\delta\phi_{,a},
\end{align}
and $\mathcal{H}_{0}\big|^{(2)}=\mathcal{H}_{g,0}\big|^{(2)}+\mathcal{H}_{f,0}\big|^{(2)}$,
\begin{align}\begin{split}\label{con20}
\mathcal{H}_{g,0}\big|^{(2)}&=a\delta\pi_{ab}\delta\pi^{ab}-\frac{1}{2}a(\delta\pi)^2+\frac{a^{-1}p}{3}\delta\pi^{ab}\delta q_{ab}-\frac{a^{-1}p}{6}\delta\pi\delta q+\frac{5a^{-3}p^2}{72}\delta q_{ab}\delta q^{ab}\\ 
&+\frac{7a^{-3}p^2}{48}(\delta q)^2+\frac{a^{-3}}{2}\left(\delta q_{ab,ab}\delta q+\frac{3}{2}\delta q_{aa,c}\delta q_{bb,c}-\delta q_{ab,b}\delta q_{ac,c}+\frac{1}{2}\delta q_{ab,c}\delta q_{ab,c}\right),\\
\mathcal{H}_{f,0}\big|^{(2)}&=K\sqrt{q}\mu^{\alpha}\left(\frac{\alpha-2}{2(\alpha-1)}\left(\frac{\delta p^{\phi}}{\bar{p}^{\phi}}\right)^2+\frac{2-\alpha}{2(\alpha-1)}\left(\frac{\delta q}{q}\right)\left(\frac{\delta p^{\phi}}{\bar{p}^{\phi}}\right)+\frac{1}{8(\alpha-1)}\left(\frac{\delta q}{q}\right)^2+\frac{1}{4}\frac{\delta q_{ab}\delta q^{ab}}{a^4}\right)\\ 
&+K\sqrt{q}\mu^{\alpha-2}\frac{\alpha}{2}\left(a^{-2}\delta^{ab}\delta\phi_{,a}\delta\phi_{,b}\right),\end{split}
\end{align}
where we lower/raise the indices with the fiducial metric $\delta_{ab}$/$\delta^{ab}$ and $\mu^{\alpha}=\left(\frac{|\bar{p}^{\phi}|}{K\alpha a^3}\right)^{\frac{\alpha}{\alpha-1}}$ is zero-order.

\section{Physical interpretation of canonical variables}\label{App1}
In the present section we give a physical (or, geometrical) interpretation to the canonical variables used throughout the paper. The geometrical quantities include expansion, shear and intrinsic curvature of the three-geometry. The material quantities include the energy density of the fluid and the flow of the fluid.

The extrinsic curvature expanded up to first order reads,
\begin{align}
K_{ab}\simeq -\frac{1}{6}ap\delta_{ab}+\frac{1}{6}a^{-1}p\delta q_{ab}-\frac{1}{12}a^{-1}p(\delta q_{ab}\delta^{ab})\delta_{ab}+a\delta\pi_{ab}-\frac{1}{2}a\delta\pi\delta_{ab},
\end{align}
where we lower/raise the indices with $\delta_{ab}$/$\delta^{ab}$.
The quantities associated with the extrinsic geometry are the expansion parameter, $\theta=K_{ab}q^{ab}$, and the shear tensor, $\sigma_{ab}=K_{ab}-\frac{1}{3}\theta q_{ab}$. When expressed in terms of the canonical variables they read
\begin{align}
\theta|^{(0)}&=-\frac{1}{2}a^{-1}p,\\
\theta|^{(1)}&=-\frac{1}{2}a^{-1}\delta\pi+\frac{1}{12}a^{-3}p(\delta q_{ab}\delta^{ab}),\\
\sigma_{ab}|^{(0)}&=0,\\
\sigma_{ab}|^{(1)}&=\frac{1}{3}a^{-1}p\left(\delta q_{ab}-\frac{1}{3}\delta q_{cd}\delta^{cd}\delta_{ab}\right)+a\left(\delta\pi_{ab}-\frac{1}{3}\delta\pi\delta_{ab}\right).
\end{align}
Let us make the Fourier transform of the first-order quantities and express them in terms of the scalar modes, i.e.
\begin{align}
\check{\theta}|^{(1)}&=-\frac{1}{2}a^{-1}\delta\pi_1+\frac{1}{4}a^{-3}p\delta q_1,\\
\check{\sigma}_{ab}|^{(1)}&=\left(\frac{1}{3}a^{-1}p\delta q_2+
\frac{3}{2}a\delta\pi^{2}\right)\left(\frac{k_ak_b}{k^2}-\frac{1}{3}\delta_{ab}\right).
\end{align}
We will denote the first-order expansion $\check{\theta}|^{(1)}$ by $\delta\check{\theta}$. We define the scalar shear $\delta{\sigma}$ as 
\begin{align}
\delta\check{\sigma}:=\frac{1}{3}a^{-1}p\delta q_2+
\frac{3}{2}a\delta\pi^{2}.
\end{align}
The quantities associated with the intrinsic geometry and their Fourier transform read
\begin{align}
\delta q&=a^4\delta q_{ab}\delta^{ab},~~\delta\check{q}= 3a^4\delta q_1,\\
\delta R&=a^{-4}(\delta q_{ab,ab}-\delta q_{aa,bb}),~~ \delta\check{R}=2a^{-4}k^2(\delta q_1-\frac{1}{3}\delta q_2),
\end{align}
whereas
\begin{align}
\delta \rho= a^{-3}a^{\frac{-3}{\alpha-1}}p^T\left(\frac{\delta p^{\phi}}{\bar{p}^{\phi}}-\frac{\delta q}{2a^{6}}\right),~~\delta\check{\rho}=a^{-3}a^{\frac{-3}{\alpha-1}}p^T\left(\frac{\delta p^{\phi}}{\bar{p}^{\phi}}-\frac{3 \delta q_1}{2a^{2}}\right).
\end{align}
We have just shown the invertible linear map between the canonical perturbation variables and the physical quantities,
\begin{align}
(\delta q_1,\delta q_2,\delta\pi_1,\delta\pi_2,\delta p^{\phi},\delta\phi)~~\longleftrightarrow~~ (\delta {q},\delta {R}, \delta{\theta},\delta{\sigma},\delta{\rho},\delta\phi).
\end{align}
Note that $\delta\phi$ has the geometrical interpretation of a quantity that determines the part of the fluid flow that is tangential to the three-surface as $\vec{U}\cdot {e}_{\bar{a}}=\mu^{-1}\delta\phi_{,a}$, where ${e}_{\bar{a}}$ are defined in Eq. (\ref{vecbasis}). The gauge-invariant quantities, given by Eqs (\ref{GIQ1}), in terms of the physical quantities read
\begin{align}
\Phi=\bar{p}^{\phi}\delta\phi+\frac{\alpha}{\alpha-1}\left(\frac{p}{4}\right) \frac{a^4}{k^2}\delta R,~~\Pi=\frac{a^3a^{\frac{3}{\alpha-1}}}{p^T}\delta\rho-\frac{3(\alpha-2)}{8}\frac{a^2}{k^2}\delta R-\frac{3(\alpha-1)}{2}a^{-2}p^{-1}\bar{p}^{\phi}\delta\phi.
\end{align}
The constraints in terms of the physical quantities read
\begin{align}
\mathcal{H}_{0}\big|^{(0)}&=-\frac{2}{3}a^3\theta^2+\frac{\alpha-1}{\alpha}a^{\frac{-3}{\alpha-1}}p^T,\\\label{gemk}
i\mathcal{H}_{\vec{k}}\big|^{(1)}&=\frac{4}{3}\left(a\delta\sigma-a^3\delta\theta\right)-\bar{p}^{\phi}\delta\phi,\\\label{gem0}
\mathcal{H}_{0}\big|^{(1)}&=\frac{2}{3}a^{2}p\delta\theta-a^3\delta R+a^3\delta\rho.
\end{align}
We may now rewrite the Dirac observables, $\Phi$ and $\Pi$, purely in terms of the geometrical quantities by making use of the above constraints,
\begin{align}
\Phi&=\frac{4}{3}(a\delta\sigma-a^3\delta\theta)+\frac{\alpha}{\alpha-1}\frac{p}{4} \frac{a^4}{k^2}\delta R,\\
\Pi&=\frac{2(\alpha-1)(\alpha-2)}{\alpha}ap^{-1}\delta\theta-2(\alpha-1)a^{-1}p^{-1}\delta\sigma+a^2\left(
\frac{6(\alpha-1)}{\alpha p^2}-\frac{3(\alpha-2)}{8k^2}\right)\delta R.\end{align}
Once gauge-fixing conditions are imposed it may be more convenient to express the gauge-invariant variables in terms of quantities other than geometrical. Nevertheless, we are going to stick to the geometrical quantities. Let us now study from the physical perspective the gauge-fixing conditions that were discussed in this work:

\subsection*{Spatially flat slicing gauge}
In the spatially flat slicing gauge one fixes $\delta R=0$ and $\delta q=0$. The gauge-invariant variables can be expressed in terms of the two extrinsic curvature quantities, $\delta\theta$ and $\delta\sigma$,
\begin{align}
\Phi=\frac{4}{3}(a\delta\sigma-a^3\delta\theta),~~\Pi=\frac{2(\alpha-1)(\alpha-2)}{\alpha}ap^{-1}\delta\theta -2(\alpha-1)a^{-1}p^{-1}\delta\sigma.
\end{align}
By making use of the constraint (\ref{gem0}) one can also substitute the expansion for the matter density as $\delta\rho=-\frac{2}{3}a^{-1}p\delta\theta$.

\subsection*{Uniform density gauge}
In the uniform density gauge one fixes $\delta \rho=0$ and $\delta q=0$. The gauge-invariant variables can be again expressed in terms of the two extrinsic curvature quantities, $\delta\theta$ and $\delta\sigma$,
\begin{align}
\Phi&=\frac{4}{3}a\delta\sigma+a^3\left(\frac{\alpha}{6(\alpha-1)}\frac{p^2}{k^2}-\frac{4}{3}\right)\delta\theta,\\\Pi&=-2(\alpha-1)a^{-1}p^{-1}\delta\sigma+ap^{-1}\left(2(\alpha-1)-\frac{1}{4}(\alpha-2)\frac{p^2}{k^2}\right)\delta\theta
\end{align}
By making use of the constraint (\ref{gem0}) one can also substitute the expansion for the Ricci curvature as $\delta R=\frac{2}{3}a^{-1}p\delta\theta$.

\subsection*{Comoving orthogonal gauge}
In the comoving orthogonal gauge one fixes $\delta \phi=0$ and $\delta\rho+\frac{\alpha}{12(\alpha-1)}\frac{p^2}{a^8}\delta q=0$. The gauge-invariant variables can be expressed in terms of $\delta q$ and $\delta R$,
\begin{align}
\Phi=\frac{\alpha}{\alpha-1}\frac{p}{4} \frac{a^4}{k^2}\delta R,~~
\Pi=-\frac{\delta q}{2a^6}-\frac{3(\alpha-2)}{8}\frac{a^2}{k^2}\delta R.
\end{align}
By making use of the constraint (\ref{gem0}) one can substitute the metric density, or the Ricci curvature, for the expansion as $\frac{2}{3}a^{2}p\delta\theta-a^3\delta R-\frac{\alpha}{12(\alpha-1)}\frac{p^2}{a^5}\delta q=0$. Furthermore, by making use of the constraint (\ref{gemk}) one can use interchangeably the expansion or the shear as $\delta \sigma=a^{2}\delta\theta$.

\subsection*{Longitudinal gauge}
In the longitudinal gauge one fixes $\delta R-\frac{2}{3}a^{-8}k^2\delta q=0$ and $\delta\sigma=0$. The gauge-invariant variables can be expressed in terms of the  two curvatures, the expansion rate $\delta\theta$ and the intrinsic curvature, $\delta R$,
\begin{align}
\Phi&=-\frac{4}{3}a^3\delta\theta+\frac{\alpha}{\alpha-1}\frac{p}{4} \frac{a^4}{k^2}\delta R,\\
\Pi&=\frac{2(\alpha-1)(\alpha-2)}{\alpha}ap^{-1}\delta\theta+a^2\left(
\frac{6(\alpha-1)}{\alpha p^2}-\frac{3(\alpha-2)}{8k^2}\right)\delta R.
\end{align}
By making use of the gauge-fixing condition $\delta R=\frac{2}{3}a^{-8}k^2\delta q$ one can also substitute the Ricci curvature for the metric density.

\subsection*{Multi-fluid case}
The above results can be straightforwardly extended to the multi-fluid case. Below we just express the Dirac observables (\ref{multiDirac}) in terms of physical quantities,
\begin{align}\begin{split}
\Phi_i&=\bar{p}^{\phi_i}\delta\phi_i+\frac{3a^3a^{\frac{-3}{\alpha_i-1}}p^{T_i}}{2pk^2}\delta R,\\
\Pi_i&=\frac{a^3}{a^{-\frac{3}{\alpha_i-1}}p^{T_i}}\delta\rho_i+\frac{3}{8}\frac{a^2}{k^{2}}\delta R.\end{split}
\end{align}
The specification of the meaning of the Dirac observables for each choice of gauge-fixing condition easily follows.

\section{Relation to the configuration space approach}\label{App2}
In the standard approach (see e.g. \cite{Malik:2008im}), the perturbed metric is expressed as
\begin{align}
\ud s^2=a^2\left[-(1+2\phi)\ud\eta^2+2B_a\ud x^a\ud\eta+(\delta_{ab}+2C_{ab})\ud x^a\ud x^b\right],
\end{align}
where $B_a$ and $C_{ab}$ admit the usual scalar-vector-tensor decomposition,
\begin{align}
B_a=B_{,a}-S_a,~~C_{ab}=-\psi\delta_{ab}+E_{,ab}-F_{(a,b)}+\frac{1}{2}h_{ab}.
\end{align} 
Notice that from the beginning it is assumed that at zero order $N=a$. They can be related to the perturbation variables used in the present paper and expressed in the momentum space as follows (we omit the inverted hats):
\begin{align}\begin{split}
&{\psi}=-\frac{3\delta q_1-\delta q_2}{6a^2},~~{E}=-\frac{\delta q_2}{2a^2k^2},~~\phi=\frac{\delta N}{a},~~B=\delta n,~~{S}_a=-k(\delta {N}^{\vec{v}}v_{a}+\delta {N}^{\vec{w}}w_{a}),\\ &F_a=\frac{i}{2\sqrt{2}}(\delta q_3v_{a}+\delta q_4w_{a}),~~{h}_{ab}=\frac{k^2}{\sqrt{2}}(v_a w_b+v_b w_a)\delta q_5+\frac{k^2}{\sqrt{2}}(v_a v_b-w_a w_b)\delta q_6.\end{split}
\end{align}
For example, in the {\it longitudinal gauge} at the canonical level we have $\delta q_2=0$, $\delta n=0$ and  $\frac{\delta N}{N}=-\frac{1}{2a^2}\delta q_1$ that leads to $E=0$, $B=0$ and $\phi={\psi}$, and hence
\begin{align}\label{gauge3}
\ud s^2=a^2\left[-(1+2\psi)\ud\eta^2+(1-2\psi)\delta_{ab}\ud x^a\ud x^b\right],
\end{align}
where $\psi=-\frac{a^{-6}}{6}\delta q$ is proportional to the density perturbation.

{
In both approaches, ours and the standard one, the scalar gauge-invariant quantities are defined as the scalar quantities which are invariant with respect to $2$ scalar-type gauge transformations. Since the (scalar) configuration space ($\phi$,$\psi$, $B$, $E$, $\delta\phi$)\footnote{Note that $\phi$ is the metric perturbation, whereas $\delta\phi$ is the fluid perturbation.} is $5$-dimensional there must be $5-2=3$ gauge-invariant quantities in the configuration space. Since the (scalar) kinematical phase space ($\delta q_1$, $\delta q_2$, $\delta\pi_1$, $\delta\pi_2$, $\delta p^{\phi}$, $\delta\phi$) is $6$-dimensional there must be $6-2=4$ gauge-invariant quantities in the kinematical phase space. Notice that the kinematical phase space in not simply twice as large as the configuration space because the lapse function and the scalar part of the shift vector (and their canonical momenta) do not belong to the kinematical ADM phase space. A standard set of gauge-invariant variables in the configuration space are the Bardeen potentials, $\Psi$ and $\Phi$\footnote{ Note that the same symbol $\Phi$ is used to denote one of the Bardeen potentials and one of the Dirac observables. It should be clear, however, which quantity is meant by it from the context in which it appears.},  which are introduced purely in terms of the gravitational variables and the velocity potential $\varphi_v^{(gi)}$ of the hydrodynamical fluid. They are defined e.g. in \cite{Mukhanov:1990me} with Eqs (3.13) and (10.42) based on earlier definitions therein. On the other hand, in the kinematical phase space we have found $4$ gauge-invariant variables, $\delta D_i$. We define them in Eqs (\ref{D1}-\ref{D4}) of the present paper. 

Because of this difference in the number of unconstrained gauge-invariant variables in both approaches it is difficult to make comparisons between them. Nevertheless, it is quite possible and desired to compare the physical degrees of freedom which are defined as gauge-invariant quantities that satisfy the constraints. There are $2$ scalar constraints: the Hamiltonian constraint and the scalar part of the vector constraints. Thus, the imposition of the constraints leads to $3-2=1$ independent  physical degree of freedom in the standard approach and $4-2=2$ independent physical degrees of freedom in our approach. The latter are called Dirac observables. Now the number of the Dirac observables is twice as large as the number of physical degrees of freedom in the standard approach. This is simply because the Dirac observables define the physical phase space whereas the physical variable of the standard approach defines the physical configuration space and the latter must be half of the former in dimensionality. Moreover, the physical configuration variable must be a linear combination of the Dirac observables. Below we give this combination explicitly. All the above considerations readily extend to many-fluid case.

\subsection*{Dirac observables and Bardeen potentials}

In single-fluid case, there is only one independent Bardeen potential when the constraints are imposed and we have $\Psi=\Phi\footnote{Here $\Phi$ denotes the Bardeen potential.}$. Making use of the longitudinal gauge and the formulas (\ref{gauge1}), (\ref{gauge2}) and (\ref{gauge3}), we find the relation between the potential $\Psi$ and  the Dirac observables (\ref{GIQ1}), $\Phi$ and $\Pi$, which in this gauge reads
\begin{align}
\Psi\big|_{longitudinal}:=\psi=-\frac{(\alpha-2)p}{16a^2k^2}\Phi\big|_{longitudinal}-\frac{\alpha p^2}{24(\alpha-1)k^2}\Pi\big|_{longitudinal}.
\end{align}
Since the perturbation variables $\Psi$, $\Phi$ and $\Pi$ are gauge-invariant the validity of this relation is not restricted to the longitudinal gauge, i.e., it must hold universally. Making use of the Hamilton equations (\ref{HEMsfs}) one shows that the so defined variable $\Psi$ satisfies the following equation
\begin{align}
\Psi''-\frac{\alpha}{2(\alpha-1)}p\Psi'+\frac{k^2}{\alpha-1}\Psi=0,
\end{align}
where $':=\frac{\partial}{\partial\eta}$. This result agrees with Eq. (12.9) of \cite{Mukhanov:1990me}.

\subsection*{Dirac observables and Mukhanov-Sasaki variable}
In the standard approach one expands to second order the action and then reduces its form by reducing the number of configurational variables to one by applying the gauge transformations and the constraint equations. The widely used physical variable is the Mukhanov-Sasaki variable, denoted by  $v$, and defined in Eq. (10.43) of \cite{Mukhanov:1990me}.  One may show that the Mukhanov-Sasaki variable and the Dirac observables are related via
\begin{align}
\Phi=-\frak{l}\sqrt{w(w+1)}~ap\cdot v,~~\frak{l}=\sqrt{\frac{8\pi G}{3}},
\end{align}
$(w=\frac{1}{\alpha-1})$. Switching to the Mukhanov-Sasaki variable $v$ and its canonical conjugate $\pi_v$ via the canonical transformation,
\begin{align}
\Phi\rightarrow v:=\frac{\Phi}{\frak{l}\sqrt{w(w+1)}~ap},~~\Pi\rightarrow \pi_v:=\frak{l}\sqrt{w(w+1)}~\left(ap\cdot \Pi-\frak{g}^{-1}\frac{3w-1}{2(w+1)w}a^{-1}\Phi\right),\end{align}
transforms the physical Hamiltonian (\ref{SSFS}),
\begin{align}
{\bf H}=N\mathcal{H}_{0}\big|^{(0)}+
N\int\left(\frak{g}\frac{w(w+1) ap^2}{12}\Pi^2 +\frak{g}^{-1}\frac{3}{w+1}\frac{a^{-3}k^2}{p^2}\Phi^2\right),~~\frak{g}=16\pi G,
\end{align}
to its more common form based on the variable $v$,
\begin{align}
{\bf H}=N\mathcal{H}_{0}\big|^{(0)}+
\frac{N}{2a}\int\left( \pi_v^2 +\left[wk^2 -U\right]v^2\right),~~U:=-\frac{3w-1}{72}p^2.
\end{align}
Making use of the background Hamilton equations we find
\begin{align}\label{potential}
U=-\frac{3w-1}{72}p^2=\frac{a''}{a}=\frac{z''}{z},~~z=\sqrt{\frac{3}{2}}\sqrt{\frac{w+1}{w}}a,
\end{align}
which is in agreement with Eq. (10.62) of \cite{Mukhanov:1990me}.
}

\end{document}